\documentclass[journal]{IEEEtran}
\pdfoutput=1 
\IEEEoverridecommandlockouts
\usepackage{cite}
\usepackage{amsmath,amssymb,amsfonts,bm}
\usepackage{algorithm}
\usepackage{algorithmic}
\usepackage{epsfig}
\usepackage{graphicx}
\usepackage{subfigure}
\usepackage{epstopdf}
\usepackage{stfloats}
\usepackage{url}
\usepackage[percent]{overpic}
\usepackage{rotating}
\graphicspath{{figures/}}
\usepackage{textcomp}
\usepackage{color,xcolor}
\usepackage[framemethod=tikz]{mdframed}
\usepackage{multirow}
\usepackage{booktabs}
\usepackage{framed}
\usepackage{color}
\usepackage[breaklinks,colorlinks,linkcolor=black,citecolor=black,urlcolor=black]{hyperref}
\usepackage[revision]{revdiff}
\newcommand{\norm}[1]{{\|{#1}\|}}
\newcommand{\norml}[1]{{\|{#1}\|}}
\newcommand{\diag}[1]{{\mathrm{diag}(\bm #1)}}
\def\BibTeX{{\rm B\kern-.05em{\sc i\kern-.025em b}\kern-.08em
T\kern-.1667em\lower.7ex\hbox{E}\kern-.125emX}}

\newtheorem{thm}{Theorem}

\begin{document}

\title{Non-Common Band SAR Interferometry {via Compressive Sensing}}
\author{Huizhang~Yang,~\IEEEmembership{Student Member,~IEEE},
	Chengzhi~Chen,
	Shengyao~Chen,~\IEEEmembership{Member,~IEEE},
    Feng~Xi,~\IEEEmembership{Member,~IEEE},
	and Zhong~Liu,~\IEEEmembership{Member,~IEEE}
	\thanks{This work was supported in part by the National Natural Science Foundation of China under Grants 61671245, 61571228. {(Corresponding author: Shengyao Chen.)}

    {H. Yang, C. Chen, S. Chen, F. Xi, and Z. Liu are with the School of Electronic and Optical Engineering, Nanjing University of Science and Technology, Nanjing, China. (e-mail:hzyang@njust.edu.cn; chenshengyao@njust.edu.cn).}
    }
}
{}
\maketitle

\begin{abstract}
	\rnew{This is the preprint version, to read the final
version please go to IEEE Transactions on Geoscience and
Remote Sensing on IEEE Xplore.} To avoid decorrelation, conventional SAR interferometry (InSAR) requires that interferometric images should have a common spectral band and the same resolution after proper pre-processing. For a high-resolution (HR) image and a low-resolution one, the interferogram quality is limited by the low-resolution one since the non-common band between two images is usually discarded. In this paper, we try to establish an InSAR method to improve interferogram quality {by means of exploiting the non-common band}.
	To this end, we first define a new interferogram, which has the same resolution as the HR image.
	Then we formulate the interferometric {relation} between the two images into a compressive sensing (CS) model, which contains the proposed HR interferogram. With the sparsity of interferogram in appropriate domains, we {model} the interferogram formation as a typical sparse recovery problem.
	Due to {the} speckle effect in coherent radar imaging, the sensing matrix of our CS model is inherently random. We theoretically prove that the sensing matrix satisfies restricted isometry property, and thus the interferogram recovery performance is guaranteed.
	Furthermore, we provide a fast interferogram formation algorithm by exploiting computationally efficient structures of the sensing matrix. {Numerical experiments show that the proposed method provide better interferogram quality in the sense of reduced phase noise and obtain extrapolated interferogram spectra with respect to common band processing.
}
\end{abstract}


\IEEEpeerreviewmaketitle

\section{Introduction}
\IEEEPARstart{S}{ynthetic} aperture radar interferometry (InSAR) exploits two or more complex-valued SAR images to derive interferometric information about an area of interest.
Conventionally, this technique requires that interferometric images should have a common spectral band, so that each resolution cell has the same scattering contribution \cite{Ferretti2007InSAR}. Hence, the essence of conventional InSAR is a \emph{common band (CB) interferometry} technique.
If the CB requirement is not satisfied, CB filtering should be applied to remove decorrelated contributions, so that interferometry can be carried out pixel by pixel \cite{Ferretti2007InSAR,Guarnieri1999Combination}. However, CB filtering decreases the resolution of the images, leading to the reduction of interferogram quality.
Thus it is meaningful to study how to reduce this kind of interferogram quality loss.

 On the other hand, InSAR also faces several challenges to acquire high resolution in SAR imaging.
For example, high resolution requires large transmitted bandwidth and high sampling/data rate, putting great pressure on sampling devices and data storage. In spaceborne SAR, it often needs to sacrifice azimuth resolution in return for wide coverage and short revisit time \cite{cumming}.
These problems suggest that low-resolution (LR) observation is also important for SAR.

\begin{figure}
	\centering
	\includegraphics[width=8.5cm]{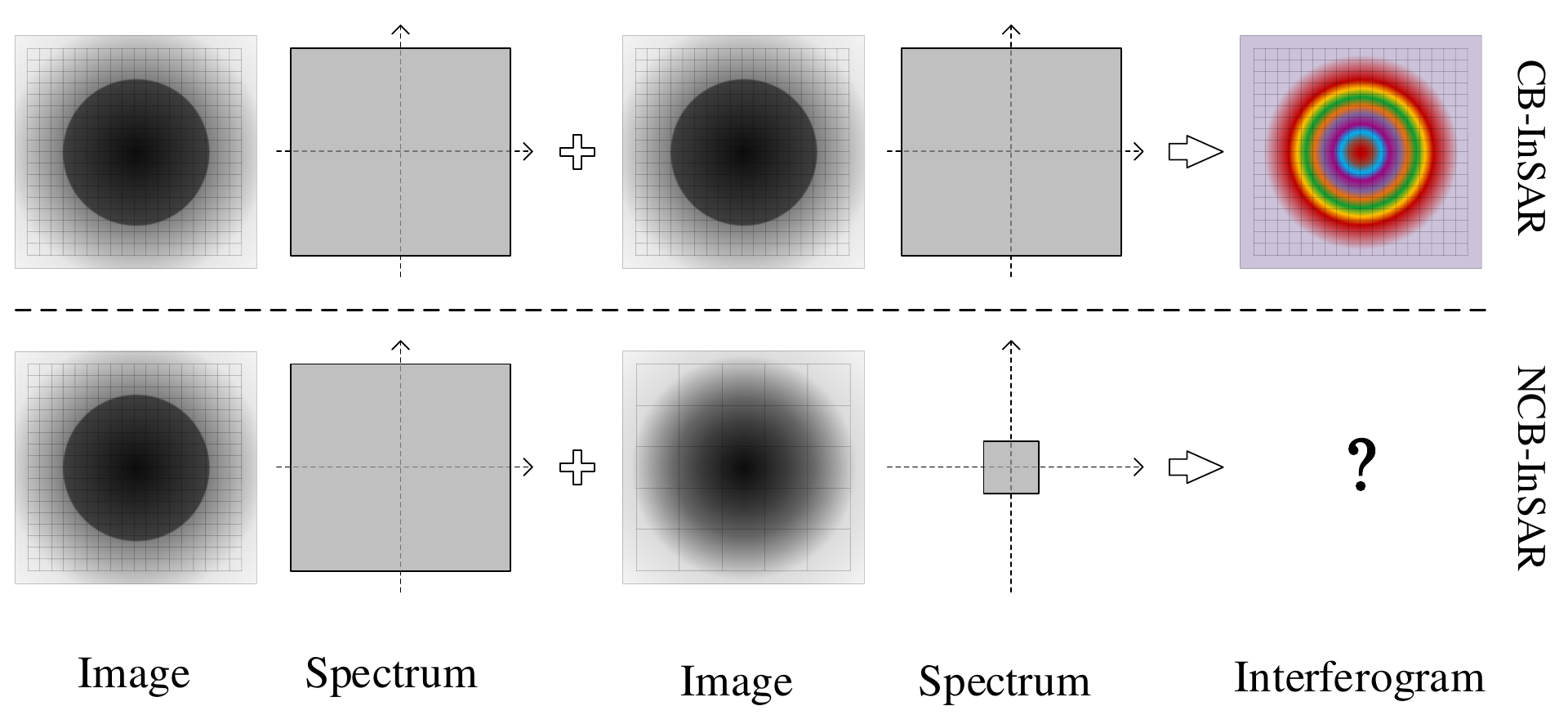}\\
	\caption{Conceptual explanations for {the} conventional CB-InSAR (top) and the proposed NCB-InSAR (bottom). The grid size{s} indicates the fineness of resolution{s}. Conventional CB-InSAR uses two images with the same resolution; {NCB-InSAR considers the interferometry between a high-resolution image and a low-resolution one}.}\label{insar_con}
\end{figure}

Motivated by these problems, this paper tries to achieve the goal of recovering a high-quality interferogram from a HR master image and a LR slave one.
To this end, we propose a novel InSAR method, which improves interferogram quality by exploiting the non-common band (NCB).
We refer to our method as \emph{NCB SAR interferometry} (NCB-InSAR).
The main difference between NCB-InSAR and conventional CB-based InSAR (CB-InSAR) is illustrated in Fig. \ref{insar_con}.
In NCB-InSAR, we first introduce a new HR {complex} interferogram $\bf u$, which has the same resolution and phase as the conventional flattened {complex} interferogram.
Then we formulate the interferogram $\bf u$ into a compressive sensing (CS) \cite{candes2008introduction} model as
\begin{equation}\label{s0e1}
	{\mathbf{y}_s}=\mathbf{A}_m \mathbf{x},
\end{equation}
where $\mathbf{y}_s$ is the LR slave image vector, $\mathbf{x}$ is the coefficient vector of the HR interferogram $\bf u$ in some basis, {and $\mathbf{A}_m$ is a sensing matrix defined by a lowpass filtering matrix $\bf Q$ and  a basis $\bf W$ as
\begin{equation}\label{}
   \mathbf{A}_m=\mathbf{Q}\,\diag{\theta_m}\,\mathbf{W}.
\end{equation}
where ${\bm\theta_m}$ is the complex exponential phase vector related to the master image.}
{In this paper we focus on spatially extended targets having homogeneous fringes, for cases like mountain and desert areas, and for a simple example, a slope}.
Then similar to natural images that share geometrical natures \cite{Mairal2008}, the interferogram in these cases is often sparse in  discrete cosine transform (DCT) or wavelet domains \cite{Xu2015}. Exploiting this sparse property, we can recover the sparse coefficients by solving an $\ell_1$ minimization problem
\begin{equation}\label{bpdn}
	\min\limits_{\mathbf{x}}^{}   \norm{{\mathbf{A}_m}\mathbf{x}-\mathbf{y}_s}_2^2+\lambda\norm{\mathbf{x}}_1,
\end{equation}
where $\lambda$ is a regularization parameter. Then {an estimate $\hat{\bf u}$ for the HR interferogram $\bf u$} can be retrieved by transforming the recovered coefficients back to image domain. Different from CB-InSAR, the proposed NCB-InSAR method uses CS to generate interferograms instead of pixel-by-pixel {conjugate multiplication}.

It is worth noting that CS has been shown to be an efficient tool for SAR. Over the past decade, several CS applications have been introduced to SAR, and brought several benefits \cite{cetin2014sparsity}.
On the one hand, CS can reduce the sampling rate on SAR data acquisition \cite{alonso2010novel,dong2014novel,Fang2013Fast,Budillon2010Three,Wang2013Compressive}.
In \cite{alonso2010novel,dong2014novel,Fang2013Fast}, the authors randomly sub-sampled SAR echoes in azimuth and range to efficiently reduce SAR data size. In \cite{Budillon2010Three}, Budillon extended random sub-sampling in elevation dimension for SAR tomography (TomoSAR).
In \cite{Wang2013Compressive}, Wang et al. exploited CS to detect sparse moving targets with low-rate measurements.
On the other hand, CS is used to achieve super resolution as a signal processing approach\cite{Zhang2009Achieving,5482210,Zhu2010Tomographic,Zhu2011Super}. For inversed SAR, Zhang et al. illustrated that CS can generate HR inversed SAR images with limited transmitted pulses \cite{Zhang2009Achieving,5482210}. In TomoSAR, Zhu et al. demonstrated that CS leads to super resolution in elevation by irregular sampling \cite{Zhu2010Tomographic,Zhu2011Super}.
In essence, our study belongs to the signal recover problem via CS. However, it is much different from above-mentioned works. {In our work, the data acquisition does not involve random or irregular sub-sampling, while they are indispensable in above-mentioned methods.}

In the framework of CS, the sensing matrix should satisfy restricted isometry property (RIP) \cite{candes2008introduction} to guarantee the recovery performance. To this end, a conventional solution is to {randomize the measurements}. An example is drawing samples at random or irregular locations in range/azimuth/elevation, such as Xampling SAR \cite{Aberman2017}, quadrature CS SAR \cite{Yang2019a} and CS TomoSAR \cite{Zhu2010Tomographic}. As for NCB-InSAR, we find that due to {the} speckle effect in coherent radar imaging, { ${\bm\theta_m}$ is a  random vector, introducing randomness into the sensing matrix $\mathbf{A}_m$}.
{With proper resolutions and sparse bases, we can prove that the sensing matrix satisfies RIP.}
Therefore, it is not necessary to adopt an extra random sub-sampling strategy,
{which makes it convenient to be applied} to existing spaceborne and airborne SAR systems.

\begin{figure}
	\centering
	\includegraphics[width=7.5cm]{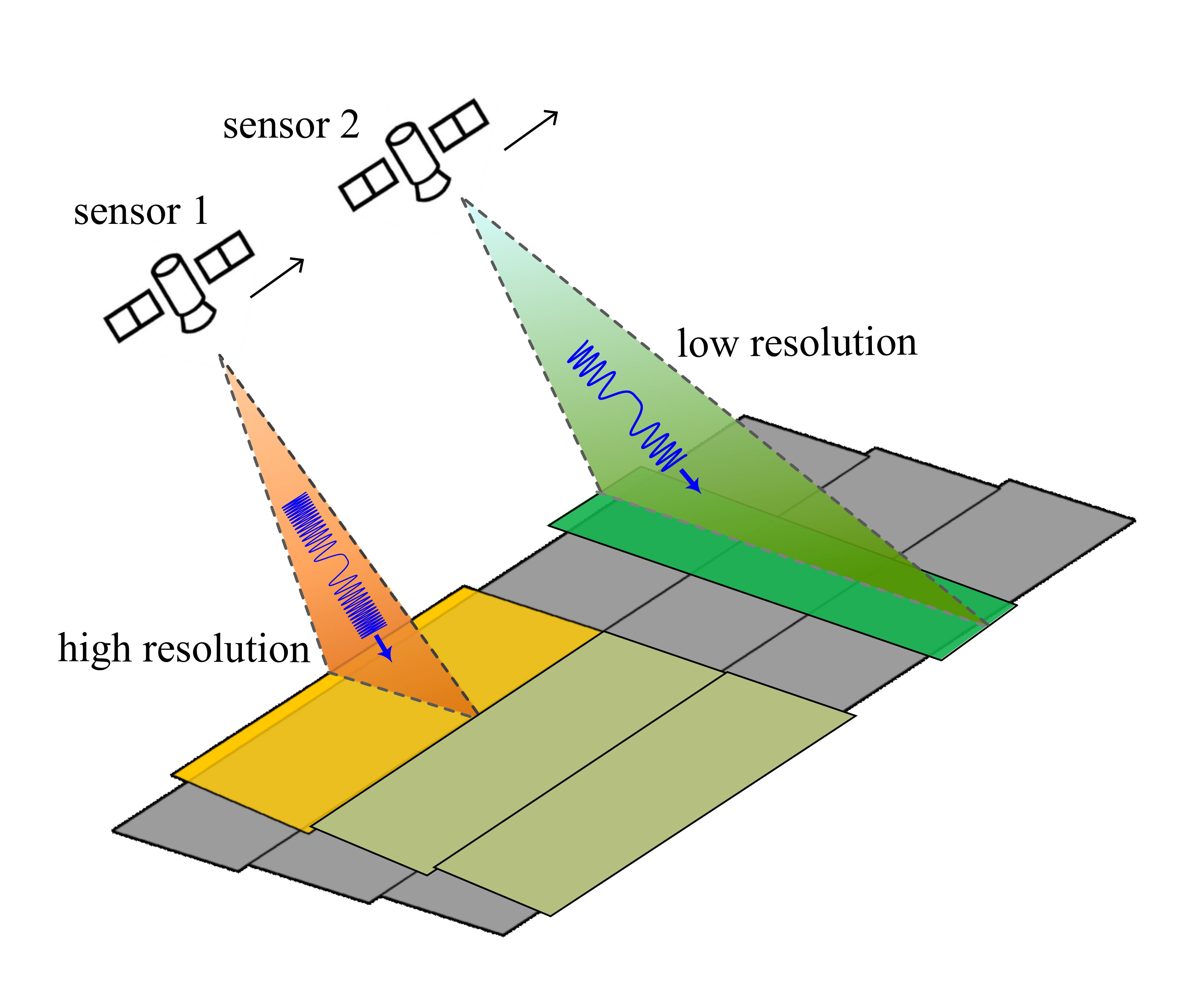}\\
	\caption{Illustration of acquisitions with different resolutions. Compared to high resolution, low resolution means small data size, wide coverage, short revisit time and small transmitted bandwidth.}\label{multimode}
\end{figure}

In practical applications, the proposed NCB-InSAR has several potential benefits. First, LR observation {results in} low data rate and size, {reducing} the cost and requirement on data acquisition and storage. Second, it {can achieve} wide coverage and short revisit time {because it allows {a} low pulse {repetition frequency}}. {It is worth pointing out that for a single HR and LR image pair, only the common footprint can be interferometrically combined, i.e., the interferometric region is limited by the scene extent of the smaller HR master image. To overcome this limitation, we can combine several HR images with one LR image, i.e., cover the LR footprint by many HR ones as explained in Fig. \ref{multimode}.} Finally, LR observation occupies small transmitted bandwidth.
For spaceborne SAR, we can combine two-satellite constellation: one sensor observes with high resolution and the other observes with low resolution, as illustrated in Fig. \ref{multimode}. This combination can be used for frequent revisiting systems to monitor changes like surface
deformation \cite{Guarnieri1999Combination}. Therefore, NCB-InSAR can provide a better way for long-term Earth observation. Besides, for multi-subband airborne SAR, the saved bandwidth can be exploited to perform other tasks \cite{Liu2017}.

The contributions of our work are {summarized} as follows:
\begin{enumerate}
  \item By means of exploiting NCB, we show that {it is possible to exploit the NCB between} a pair of HR and LR interferometric images. We formulate the interferometric relation between the two images into a CS model, which explicitly contains a HR interferogram.
  \item We prove that, owing to the speckle effect, the sensing matrix of the CS model is inherently random and satisfies RIP. Therefore, {the recovery of the interferogram is theoretically feasible and robust.}
  \item {Taking advantage of} the computationally favourable structure of the sensing matrix, we propose an efficient first-order interferogram formation algorithm, which is named as NCB InSAR via $\ell_1$ Minimization (NIL1M). Then the interferogram formation problem (\ref{bpdn}) with large scales can be solved efficiently.
  \item Based on numerical experiments using simulated data and {half-real} datasets, we find that the proposed method achieves better interferogram quality for distributed scenes having homogeneous interferometric fringes.
\end{enumerate}

The rest of this paper is organized as follows.
Section \uppercase\expandafter{\romannumeral2} reviews some basic knowledge about SAR and InSAR.
Section \uppercase\expandafter{\romannumeral3} presents the principle of NCB-InSAR.
Section \uppercase\expandafter{\romannumeral4} discusses the reconstructability and robustness of NCB-InSAR via RIP analysis.
Section \uppercase\expandafter{\romannumeral5} presents an computationally efficient interferogram formation algorithm.
Section \uppercase\expandafter{\romannumeral6} provides several supporting experiments.
Section \uppercase\expandafter{\romannumeral7} concludes this paper. Appendices A, B and C contain theoretical proofs and derivation.

\subsection*{Notation}
A continuous function and its discrete samples are denoted as $z(\tau,\eta)$ and $z[n,l]$, respectively.
The Fourier transform of $z(\tau,\eta)$ is denoted by $Z(\nu,\xi)$.
Operators are denoted by calligraphic letters, e.g., $\mathcal{F}$. The set of complex numbers is denoted by $\mathbb{C}$.
Matrices and vectors are denoted by boldface uppercase and lowercase letters, e.g., $\mathbf{A}$ and $\bf x$, respectively.

For a matrix $\bf{A}$, $\mathrm{vec} \left( \bf{A}\right)$ is the column-wise vector of $\bf{A}$, and $\text{diag}({\mathbf{A}})$ is the diagonal matrix with $\mathrm{vec}\left( {\mathbf{A}} \right)$ as its diagonal elements.
$\langle{\mathbf{A}},\mathbf{B}\rangle$ denotes the inner product of $\mathrm{vec}({\mathbf{A}})$ and $\mathrm{vec}(\mathbf{B})$.
{The matrix  $\mathbf{F}_{N}$ denotes a normalized discrete Fourier transform (DFT) matrix with size $N\times N$, which is defined as $F_N[l,k]=\frac{1}{\sqrt{N}}e^{-j\frac{2\pi}{N}(l-1)(k-1)}$ for $l,k\in\{1,\cdots,N\}$.}
The conjugate, transpose and adjoint of a matrix are denoted by ${\overline{{\left( \cdot  \right)}}}$, ${{\left( \cdot  \right)}^{T}}$ and ${{\left( \cdot  \right)}^{*}}$, respectively.
The adjoint of an operator is also denoted by ${{\left( \cdot  \right)}^{*}}$.
The ${{\ell }_{p}}$-norm of vectors is denoted by ${{\left\| \cdot  \right\|}_{p}}$.
The spectral norm and Frobenius norm of matrices are denoted by ${{\left\| \cdot  \right\|}_{2}}$ and ${{\left\| \cdot  \right\|}_{F}}$, respectively.
The symbols $\otimes $ and $\circ $ denote Kronecker product and Hadamard product, respectively.
The support {set} of a vector $\mathbf{x}$ is denoted by $\mathrm{supp}(\mathbf{x})$.

\section{Background}
This section is to review background knowledge about SAR imaging and interferometry. For simplicity, we restrict our topic to conventional stripmap InSAR.

\subsection{Bandwidth and Resolution in SAR Imaging}

SAR imaging can be viewed as the estimation of a two dimensional (2-D) backscattering field $z(\tau,\eta)$ for an area of interest, where $\tau$ and $\eta$ denote the range time and azimuth time, respectively.
To this end, a probing signal with bandwidth $B_{\tau}\times B_{\eta}$ is used to illuminate the area. The echo signal, which contains information about the illuminated area, is then observed by the sensor. After certain steps of matched filtering-based processing, the focused SAR signal can be described as \cite{Bamler1999Synthetic,Li1990Studies,soumekh1999synthetic}
\begin{equation}\label{s1e6}
	z_{mf}(\tau,\eta)= \iint z(\tau-u,\eta-v)\mathrm{sinc}\left(B_{\tau}u\right)\mathrm{sinc}\left(B_{\eta}v\right)du dv,
\end{equation}
where the sinc kernel is the impulse response function (IRF) due to bandlimited observation.

{Let} the range and azimuth frequencies {be} $\nu$ {and} $\xi$, {and} the Fourier transforms of $z(\tau,\eta)$ and $z_{mf}(\tau,\eta)$ {be} $Z(\nu,\xi)$ and $Z_{mf}(\nu,\xi)$, respectively{. We} can express (\ref{s1e6}) in frequency domain as
\begin{equation}\label{mf}
	\begin{split}
		Z_{mf}(\nu ,\xi )
		=\sqrt{\frac{1}{B_{\tau}B_{\eta}}}\mathrm{Rect}\left(\frac{\nu}{B_{\tau}},\frac{\xi}{B_{\eta}}\right)Z(\nu,\xi),
	\end{split}
\end{equation}
where $\mathrm{Rect}(\cdot)$ is a rectangular window for  bounding the spectrum   $\nu\in[-B_{\tau}/2,B_{\tau}/2]$ and $\xi \in[-B_{\eta}/2,B_{\eta}/2]$. This bandwidth gives the resolutions $1/B_{\tau}$ and $ 1/B_{\eta}$ in range-time and azimuth-time dimensions, respectively.

It should be pointed out that the rectangular spectral support in (\ref{mf}) involves assumptions of zero squint angle and constant antenna weighting. In fact, the observed spectrum is more general, which is related to the squint angle and antenna pattern. The readers may refer to Appendix 8A in \cite{cumming} for more discussions. For simplicity and without loss of generality, we consider the rectangular spectral support in this paper.

\subsection{SAR Interferometry}
The phase of a SAR image pixel is
proportional to its slant range plus a  scattering phase, i.e.,
\begin{equation}\label{}
  \phi=-2k_0R+\phi_{scat}
\end{equation}
where $R$ is the slant range, $k_0=2\pi/\lambda_0$ is the wavenumber and $\lambda_0$ is wavelength of transmitted pulse  \cite{Bamler1999Synthetic}.
Topographic elevation affects $R$, so  height information is related to the phase.

{To retrieve the height information, InSAR exploits the phase difference between two complex-valued SAR images. This involves a master image $z_m[n,l]$ and a slave image $z_s[n,l]$, which are observed at slightly different positions. The phases of each pixel pair can be expressed as
\begin{align}\label{}
      &\phi_m[n,l]=-2k_0R_m[n,l]+\phi_{scat,m}[n,l],\\ &\phi_s[n,l]=-2k_0R_s[n,l]+\phi_{scat,s}[n,l].
\end{align}
where $R_m[n,l]$, $R_s[n,l]$ are master and slave pixel ranges, and $\phi_{scat,m}[n,l]$, $\phi_{scat,s}[n,l]$ are their scattering phases, respectively.
After coregistration,  a interferometric phase can be obtained by pixel-by-pixel conjugate multiplication,
\begin{equation}\label{ifg0}
  \phi_m[n,l]-\phi_s[n,l]=-2k_0\Delta R[n,l]+\phi_{noise}[n,l],
\end{equation}
where $\Delta R[n,l]=R_m[n,l]-R_s[n,l]$, and $\phi_{noise}[n,l]=\phi_{scat,m}[n,l]-\phi_{scat,s}[n,l]$.  Note that in an ideal case we can assume that the scattering phases $\phi_{scat,m}[n,l]$ and $\phi_{scat,s}[n,l]$ are the same. But in practice, decorrelation factors such as view angle and thermal noise will corrupt this ideal assumption and consequently the phase noise $\phi_{noise}[n,l]$ occurs in the interferometric phase.}

The first term in the right side of (\ref{ifg0}) can be decomposed into a flat-Earth phase and topographic phase as
\begin{equation}\label{ifg1}
 -2k_0\Delta R[n,l]=\phi_{flat}[n,l]+\phi_{topo}[n,l],
\end{equation}
 By filtering to suppress phase noise and removal of the flat-Earth phase,  the topographic phase $\phi_{topo}[n,l]$ can be retrieved. It is notable that for repeat-pass InSAR measurements there exist residual phases due to possible atmosphere phase screen and ground deformation. In this case, $\phi_{topo}[n,l]$ should contain these residual phase components.

In the interferometric operation (\ref{ifg0}), the images are required to have the same {resolution or spectral band}. If this requirement is not satisfied, it will introduce decorrelation for each pixel pair {due to} different scattering contributions, as seen from (\ref{s1e6}). Then according to the conventional InSAR principle, CB filtering should be applied to remove the NCB to {avoid} decorrelation \cite{Guarnieri1999Combination,Ferretti2007InSAR}. However, this processing reduces the image resolution, and thus leads to {the} loss of interferogram quality \cite{Guarnieri1999Combination,Ferretti2007InSAR}.

\section{NCB-InSAR }
To avoid the above-mentioned drawback of {the} conventional CB-InSAR, this section presents our NCB-InSAR method.

\subsection{NCB Interferometric Relation}

\begin{figure}[t]
	\centering
	\includegraphics[width=6.5cm]{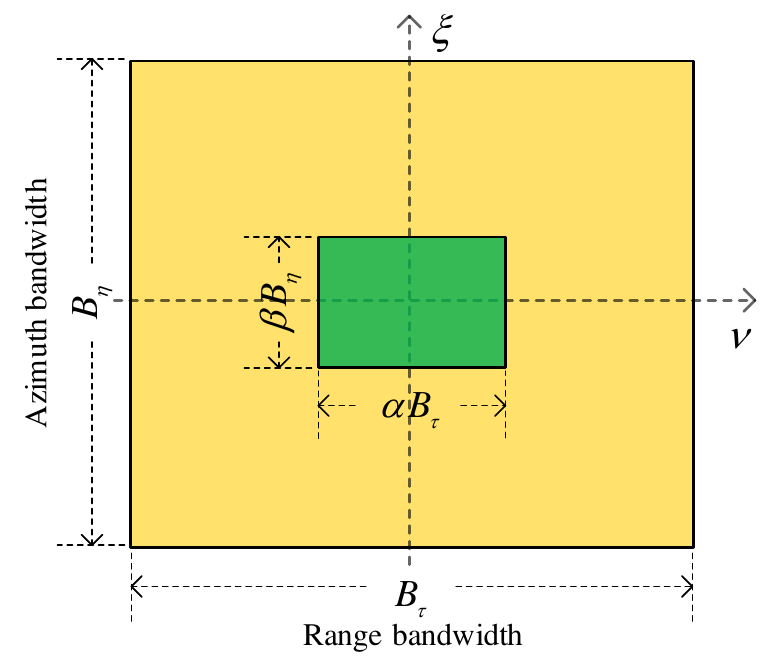}\\
	\caption{Spectral relation of the NCB InSAR image pair. The master image has a large bandwidth, and the slave image has a small bandwidth.  This setting results in two different resolutions, which is characterized by the resolution ratio $\alpha\times\beta$. The support of overlapped region is the CB, and the rest area is the NCB. This paper aims at making full use of the NCB to improve interferogram quality.}\label{dual_bandwidth}
\end{figure}

We consider the case: the master image has the bandwidth $B_{\tau}\times B_{\eta}$, while the slave image has a smaller bandwidth $\alpha B_{\tau}\times \beta B_{\eta}$  for $\alpha,\beta\in(0,1)$, as illustrated in Fig. \ref{dual_bandwidth}.
Then compared {with} the master image, the slave image has a coarser resolution. We refer to this pair of master and slave images as {the} NCB InSAR image pair. {The parameters $\alpha$ and $\beta$ are referred to as \emph{resolution ratios} in range and azimuth, respectively.}

To distinguish from the HR slave image $z_s[n,l]$ with bandwidth $B_{\tau}\times B_{\eta}$, we denote the LR slave image by $y_s[i,j]$.
According to (\ref{mf}), we can express the spectral relation between the HR and LR slave images as
\begin{equation}\label{s1e8}
	\begin{split}
		Y_{s,mf}(\nu,\xi)&=\frac{1}{\sqrt{\alpha\beta B_{\tau}B_{\eta}}}\mathrm{Rect}\left(\frac{\nu}{\alpha B_{\tau}},\frac{\xi}{\beta B_{\eta}}\right)Z_s(\nu,\xi)\\
		&=\frac{1}{\sqrt{\alpha\beta}}\mathrm{Rect}\left(\frac{\nu}{\alpha B_{\tau}},\frac{\xi}{\beta B_{\eta}}\right)Z_{s,mf}(\nu,\xi),
	\end{split}
\end{equation}
where $Z_s(\nu,\xi)$ denotes the slave spectrum to be observed, $Y_{s,mf}(\nu,\xi)$ and $Z_{s,mf}(\nu,\xi)$ denote the observed spectra with bandwidth $\alpha B_{\tau}\times\beta B_{\eta}$ and $ B_{\tau}\times B_{\eta}$, respectively.

Then in discrete domain, the 2-D discrete Fourier transforms (DFTs) of $y_s[i,j]$ and $z_s[n,l]$ are related as
\begin{equation}\label{dftrelation}
	\mathcal{F} y_s[i,j]=\frac{1}{\sqrt{\alpha\beta}}\mathcal{S}_{\alpha,\beta}\mathcal{F} z_{s}[n,l],
\end{equation}
where the operator $\mathcal{F}$ denotes 2-D DFT, and $\mathcal{S}_{\alpha,\beta}$ is a CB filter to extract the lowpass $\alpha \times \beta $ (in percentage) common {spectral band}. In this paper, DFTs are normalized.

\begin{figure*}[t]
	\centering
	\subfigure{\begin{overpic}[width=4.7cm]{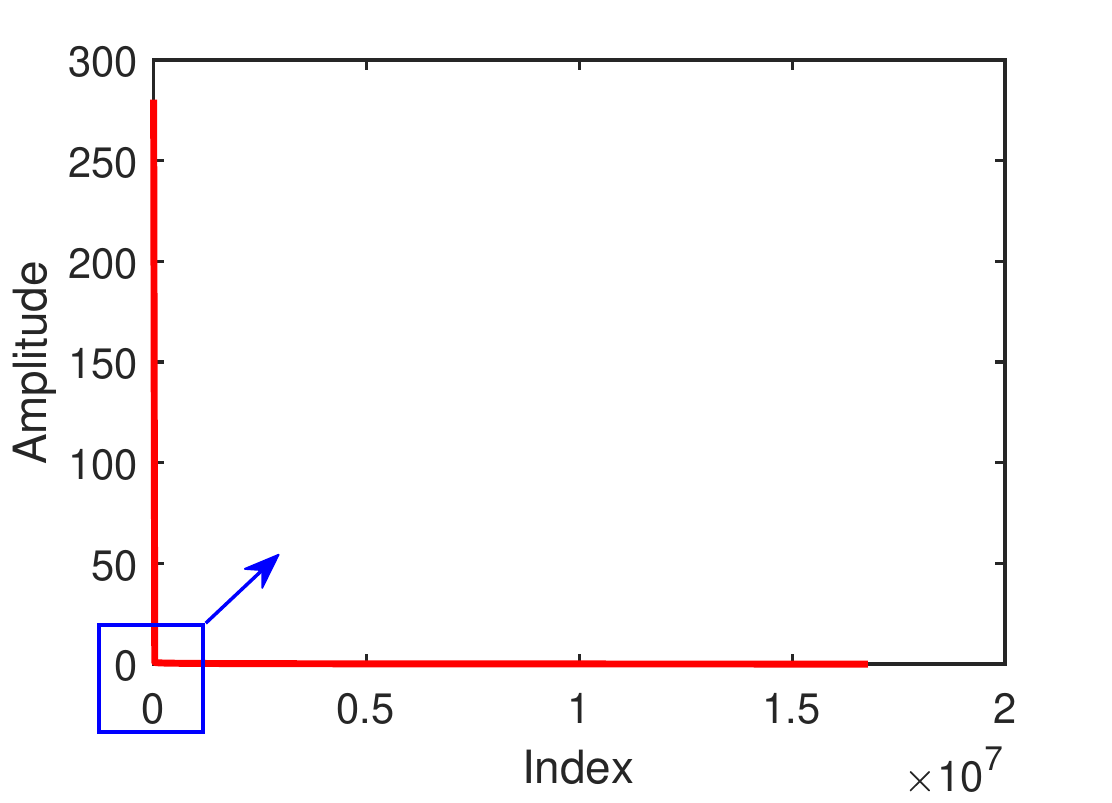}
			\put(30,20){\includegraphics[scale=0.4]{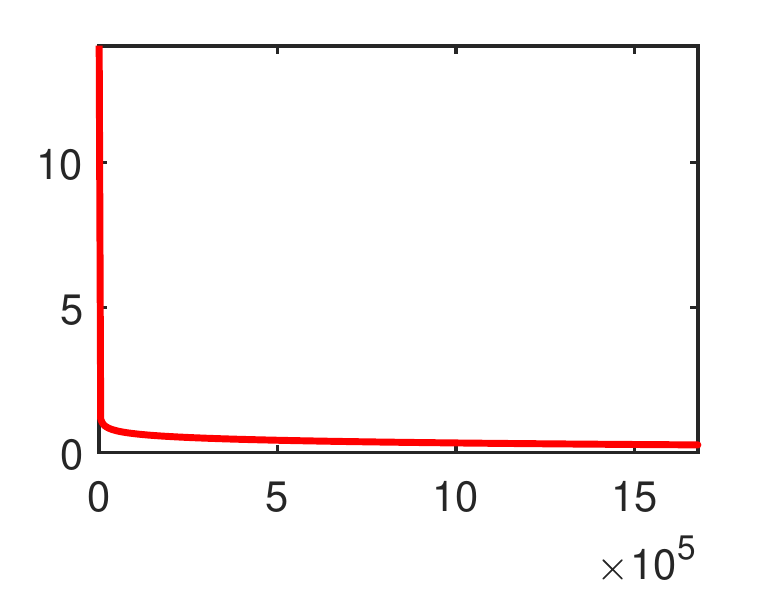}}
		\end{overpic}}\hspace{3em}
	\subfigure{
		\includegraphics[width=3.4cm]{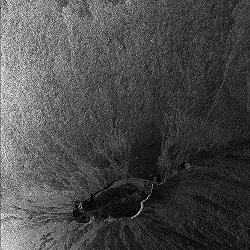}}
	\subfigure{
		\includegraphics[width=3.4cm]{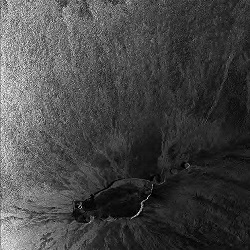}}
	\subfigure{
		\includegraphics[width=3.4cm]{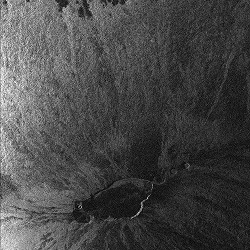}}\\
	\setcounter{subfigure}{0}\subfigure[DB-4 (top) and DCT (bottom) coefficients.]{\begin{overpic}[width=4.7cm]{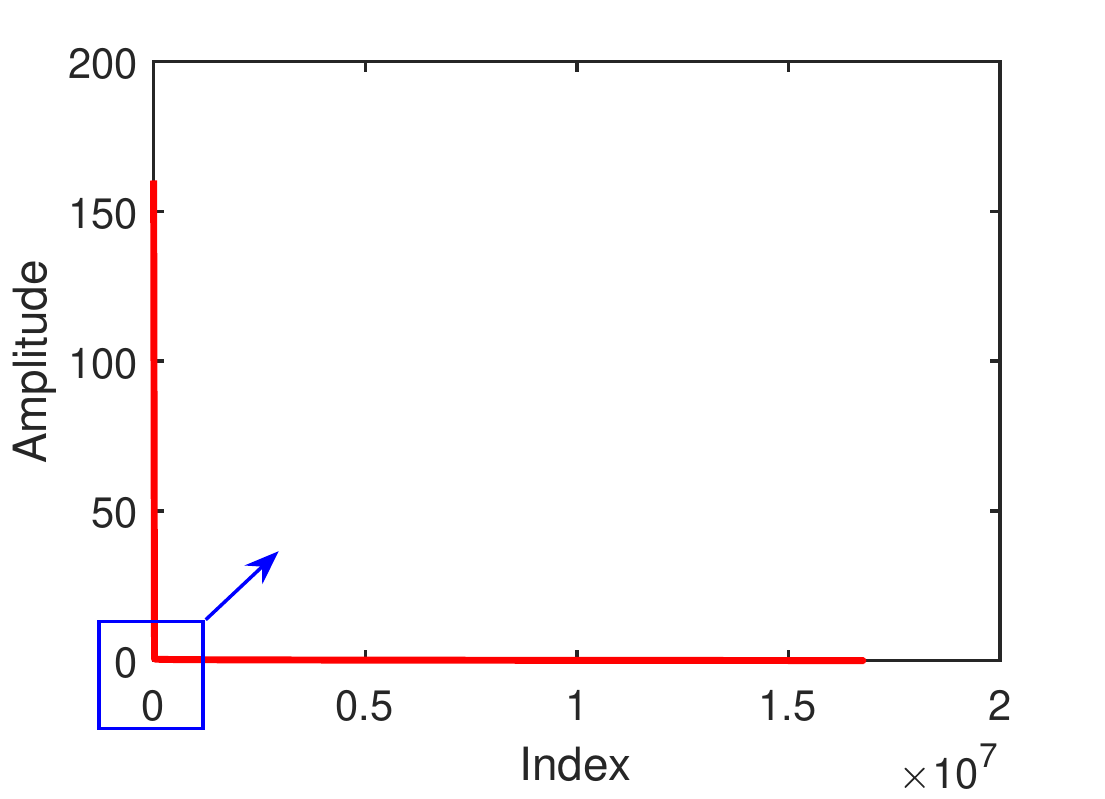}
			\put(30,20){\includegraphics[scale=0.4]{main//hawaii_wave_small-eps-converted-to.pdf}}
		\end{overpic}}\hspace{3em}
	\subfigure[Original interferogram.]{
		\includegraphics[width=3.4cm]{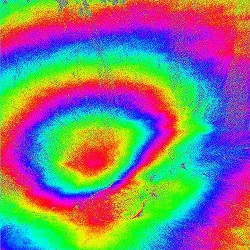}}
	\subfigure[Wavelet approximation.]{
		\includegraphics[width=3.4cm]{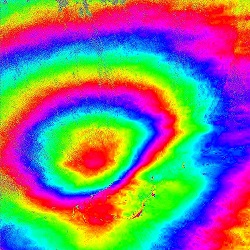}}
	\subfigure[ DCT approximation.]{
		\includegraphics[width=3.4cm]{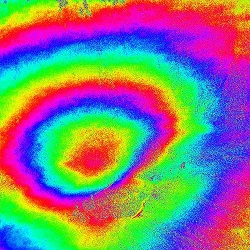}}
	\caption{ Daubechies-4 (DB-4) wavelet and DCT coefficients of a UAV SAR interferogram, and wavelet, DCT approximation of the interferogram using $5\%$ largest coefficients. The absolute value of the coefficients are sorted in descending order. The amplitude and phase of the interferograms are shown in the first and second rows in (b), (c) and (d), respectively. These figures suggests that SAR interferogram can be sparsely represented using wavelet transform or DCT.  Courtesy of NASA/JPL-Caltech.}\label{spa_coef}
\end{figure*}

According to (\ref{ifg0}) and (\ref{ifg1}), the slave image phase can be written as
\begin{equation}\label{}
	\phi_s[n,l]= \phi_m[n,l]-\phi_{flat}[n,l]-\phi_{topo}[n,l]-\phi_{noise}[n,l].
\end{equation}
Using this phase relation, we decompose the HR slave image as
\begin{equation}\label{ms}
	z_s[n,l]=\theta_m[n,l]u[n,l]+e[n,l],
\end{equation}
where
  \begin{gather}\label{}
     \theta_m[n,l]=e^{j\phi_{m}[n,l]-j\phi_{flat}[n,l]},\\
    u[n,l]=|z_s[n,l]|e^{-j\phi_{topo}[n,l]},\\
    e[n,l]=\theta_m[n,l]u[n,l](e^{-\phi_{noise}[n,l]}-1).
  \end{gather}
{In the above equations, $\theta_m[n,l]$ contains the master image phase and the flat-Earth phase (which can be computed according to a reference surface \cite{Moreira2013A});  $u[n,l]$ contains the slave image amplitude and the desired topographic phase;  $e[n,l]$  is used to model the noise introduced by $\phi_{noise}[n,l]$.} In this paper, we define $u[n,l]$ as an interferogram. This interferogram has a high resolution which equals the resolution of the HR image $z_m[n,l]$.

Inserting (\ref{ms}) into (\ref{dftrelation}), we have
\begin{equation}\label{msdft}
	 y_s[i,j]=\frac{1}{\sqrt{\alpha\beta}}\mathcal{F}^{-1}\mathcal{S}_{\alpha,\beta}\mathcal{F} \big\{\theta_m[n,l]u[n,l]\big\}+\hat{e}[n,l],
\end{equation}
{where $\hat{e}[n,l]=\frac{1}{\sqrt{\alpha\beta}}\mathcal{F}^{-1}\mathcal{S}_{\alpha,\beta}\mathcal{F} \{e[n,l]\}$}.
This linear model describes the interferometric relation of {the} NCB-InSAR image pair.

\subsection{Interferogram Formation via CS}
For ease of description, we rewrite (\ref{msdft}) in a matrix-vector form.
Firstly, we denote the sample lengths of the indices $i$, $j$, $n$ and $l$ as $I$, $J$, $N$ and $L$, respectively.
Secondly, we define two matrices, $\mathbf{\Omega}_{\alpha}$ and $\mathbf{\Omega}_{\beta}$, as row submatrices of $N\times N$ and $L\times L$ identity matrices, which extract the lowpass $\alpha\times\beta$ image spectrum in range and azimuth, respectively.
Finally, we define ${\mathbf{Y}_s}$, $\mathbf{\Theta}_m$ $\mathbf{U}$ and {$\widehat{\mathbf{E}}$} as the matrix forms of $y_s[i,j]$, $\theta_m[n,l]$, $u[n,l]$ and {$\hat{e}[n,l]$}, respectively.
Using these notations, we rewrite (\ref{msdft}) in a matrix form as
\begin{equation}\label{s3e2}
	{\mathbf{Y}_s}=\frac{1}{\sqrt{\alpha\beta}}\mathbf{F}_{I}^*\mathbf{\Omega}_{\alpha}\mathbf{F}_{N} (\mathbf{\Theta}_m\circ\mathbf{U}) \mathbf{F}_{L}^T\mathbf{\Omega}_{\beta}^T\overline{\mathbf{F}}_{J}+\widehat{\mathbf{E}}.
\end{equation}
Denoting ${\mathbf{y}_s}=\mathrm{vec}({\mathbf{Y}_s})$ and ${\mathbf{u}}=\mathrm{vec}({\mathbf{U}})$, (\ref{s3e2}) can be expressed in a matrix-vector form as
\begin{equation}\label{zmu}
	\mathbf{y}_s=\mathbf{H}_m\mathbf{u}+\hat{\mathbf{e}},
\end{equation}
where $\hat{\mathbf{e}}=\mathrm{vec}(\widehat{\mathbf{E}})$ and
\begin{equation}\label{mesmtx}
\begin{split}
	\mathbf{H}_m&=\frac{1}{\sqrt{\alpha\beta}}(\mathbf{F}_{J}^* \mathbf{\Omega}_{\beta}{\mathbf{F}}_{L}\otimes\mathbf{F}_{I}^*\mathbf{\Omega}_{\alpha}\mathbf{F}_{N} ) \mathrm{diag}(\mathbf{\Theta}_m)\\
&=\mathbf{Q}\diag{\theta_m}
\end{split}
\end{equation}
is an $IJ\times NL$ matrix. {Here $\bf Q$ denotes the lowpass filtering in the right side of (\ref{mesmtx}), which can also be viewed as a sinc blurring kernel, and ${\bm\theta}_m=\mathrm{vec}(\mathbf{\Theta}_m)$. The subscript ``$m$'' indicates that $\mathbf{H}_m$ is related to the master image. Because this matrix seems to perform measurement on $\bf u$, we refer to it as the measurement matrix.
{Since $IJ<NL$}, (\ref{zmu}) is under-determined.
Therefore, unlike conventional CB-InSAR, interferogram formation from the NCB-InSAR image pair is an ill-posed problem, and thus it cannot be solved directly.

Our {solution} is to use a CS approach to solve this ill-posed problem. According to CS theory, if the interferogram $\bf u$ is sparse in  {appropriate} bases, {it is possible to successfully recover it}.
{As we know, wavelet} transform and discrete cosine transform (DCT) are {efficient} for sparse representation in image processing{,} and {have been widely used to sparsely represent natural images} \cite{Jansen2001Noise}. {Similarly}, complex-valued SAR interferograms also exhibit sparsity in wavelet or DCT domain {\cite{Xu2015}}. For example, as the UAV SAR interferogram shown in Fig. \ref{spa_coef}(b),  its most wavelet and DCT coefficients, shown in Fig. \ref{spa_coef}(a), are small. In fact, only  $5\%$ largest coefficients can well approximate the original interferogram, as shown in Fig. \ref{spa_coef}(c) and (d).

Based on above discussions, we express the interferogram $\bf{u}$ by a basis $\bf{W}$ and a sparse vector $\bf x$ as
\begin{equation}\label{uWx}
\mathbf{u}=\mathbf{W x}.
\end{equation}
Inserting (\ref{uWx}) into (\ref{zmu}) and defining a sensing matrix
\begin{equation}\label{sensmtx}
	{\mathbf{A}_m}=\mathbf{H}_m\mathbf{W},
\end{equation}
we translate (\ref{zmu}) into a CS model as
\begin{equation}\label{s3e3}
	{\mathbf{y}_s}=\mathbf{A}_m \mathbf{x}+\hat{\mathbf{e}}.
\end{equation}
According to CS, if the sensing matrix {$\mathbf{A}_m$} satisfies RIP, we can recover the interferogram $\bf{u}$ by solving the $\ell_1$ minimization problem
\begin{equation}\label{socp}
	\min\limits_{\mathbf{x}}\norm{\mathbf{x}}_1  \quad
	\text{s.t. }              \left\|{\mathbf{A}_m}\mathbf{x}-\mathbf{y}_s\right\|_2\le\epsilon,
\end{equation}
where $\epsilon$ is a relaxation parameter accounting for noise factors. In this paper, we consider the Lagrangian form of (\ref{socp}), i.e., the problem (\ref{bpdn}). Now, this problem can be solved by a variety of algorithms, such as the truncated Newton interior-point method \cite{l1ls} and first-order algorithms \cite{nesterov,donoho2009Message,RN23,Becker2011}. With the recovered coefficient $\bf{x}$, we can then reconstruct the interferogram by (\ref{uWx}).

\section{RIP Analysis}
As we know, to ensure recovery performance in the framework of CS, the sensing matrix should satisfy RIP or incoherence \cite{candes2008introduction}.
On the other hand, the sparse representation may be mismodeled when the interferogram is not sparse enough.
Besides, InSAR observations are usually corrupted by various noise sources, such as thermal noise, quantization, atmosphere and look angle.
So, it is desired that the proposed NCB-InSAR {is theoretically reconstructable and robust} in the presence of noise and mismodeling.
In CS, RIP is a widely used {condition} for analyzing both two requirements. In this section, we therefore discuss the RIP of  NCB-InSAR.
{In the following discussions, the basis $\bf W$ is assumed to be orthonormal.}

\subsection{RIP and Robust Recovery}
The RIP defines a sequence of restricted isometry constants (RICs):
the $k$-th RIC of a matrix $\mathbf{A}$ is the smallest constant $\delta _k\in \left( 0,1 \right)$ satisfying
\begin{equation}\label{s3}
	\left( {1 - {\delta _k}} \right)\left\| {{\bf{ x}}} \right\|_2^2 \leqslant \left\| {{{ \mathbf{A} \mathbf{x}}}} \right\|_2^2 \leqslant \left( {1 + {\delta _k}} \right)\left\| {{\bf{ x}}} \right\|_2^2,
\end{equation}
for all $k$-sparse vectors $\mathbf{{x}}$.
The matrix $\mathbf{A}$ is said to satisfy $k$-order RIP if the RIC $\delta _{2k}$ is remarkably less than $1$ \cite{Foucart2013A}.

The RIC can provide the upper bound of recovery error in (\ref{socp}) for sparse and {compressible} signals.
Specifically, if ${{\delta }_{2k}}<\sqrt{2}-1$, the recovery error between the sparse or {compressible} vector $\mathbf{{x}}$ and the solution ${{\mathbf{{x}}}^{\#}}$ of (\ref{socp}) will obey  \cite{Cand2008The}
\begin{equation}\label{s4}
	{{\left\| \mathbf{{x}}-{{{\mathbf{{x}}}}^{\#}} \right\|}_{2}}\le {{C}_{1}}\frac{{{\sigma }_{k}}{{\left( {\mathbf{{x}}} \right)}_{1}}}{\sqrt{k}}+{{C}_{2}}\epsilon.
\end{equation}
In (\ref{s4}), ${{\sigma }_{k}}{{\left( {\mathbf{{x}}} \right)}_{1}}={{\inf }_{\mathrm{supp}(\mathbf{v})\le k}}{{\left\| \mathbf{{x}}-\mathbf{{v}} \right\|}_{1}}$ denotes the best $k$-term approximation error of $\mathbf{{x}}$ in ${{\ell }_{1}}$ norm, and ${{C}_{1}}$ and ${{C}_{2}}$ are positive constants depending on ${\delta _{2k}}$ \cite{Foucart2013A}.
Briefly speaking, the smaller the RIC ${\delta _{2k}}$ is, the smaller the constant $C_1$, $C_2$ and the best $k$-term approximation error will be.
So, if the sensing matrix $\mathbf{A}_m$ satisfies RIP, the recovery error ${{\| \mathbf{{x}}-{{{\mathbf{{x}}}}^{\#}} \|}_{2}}$ is well bounded, i.e., the recovery is guaranteed and is robust in the presence of noise and mismodeling.

In CS, random matrices take the leading role in the design of sensing matrices satisfying RIP, such as Gaussian matrix, partial Fourier matrix and random circulant matrix \cite{Foucart2013A}. In the literature, various RIP conditions have been established for a variety of random matrices from the probabilistic perspective. Besides the optimal Gaussian matrix, many other types of random matrices that satisfy suboptimal or quadratic RIP conditions also provide guaranteed recovery performance, such as the random circulant matrix \cite{Romberg2008Compressive,rauhut2009circulant,Haupt2010Toeplitz}.
In the following, we will also provide a quadratic RIP for the  sensing matrix $\mathbf{A}_m$ of NCB-InSAR.

\subsection{Speckle Effect and Random Sensing Matrix}
From the CS model of NCB-InSAR, we find that the sensing matrix $\mathbf{A}_m$ is inherently random. {Consider distributed scatterers as in Fig. \ref{speckle}.} Due to the {speckle effect} in SAR imaging, each pixel of a SAR image is the superposition of many independent elementary scatterers, as explained in Fig. \ref{speckle}. According to the central limit theorem, the phase of each pixel is independent random variables uniformly distributed in $[-\pi,\pi]$ \cite{Ferretti2007InSAR}. Hence, the phase matrix $\mathbf{\Theta}_m$ is a random matrix with i.i.d. entries obeying uniform distribution in the unit circle in $\mathbb{C}$. This random phase can be explicitly seen from the example in Fig. \ref{ap}. Consequently, the phase matrix $\mathbf{\Theta}_m$ injects randomness into the sensing matrix $\mathbf{A}_m$ via (\ref{mesmtx}), and then $\mathbf{A}_m$ is also random.
It is notable that, different from sensing matrices in other CS applications, $\mathbf{A}_m$ exploits the inherent randomness of speckle effect instead of artificially injecting randomness. Therefore, the proposed NCB-InSAR do not need to adopt an extra random sub-sampling strategy.
Besides, $\mathbf{A}_m$ is a structured matrix, which is different from Gaussian matrices.
Next, we will prove that this structured matrix also satisfies RIP.

\begin{figure}
	\centering
	\includegraphics[width=7cm]{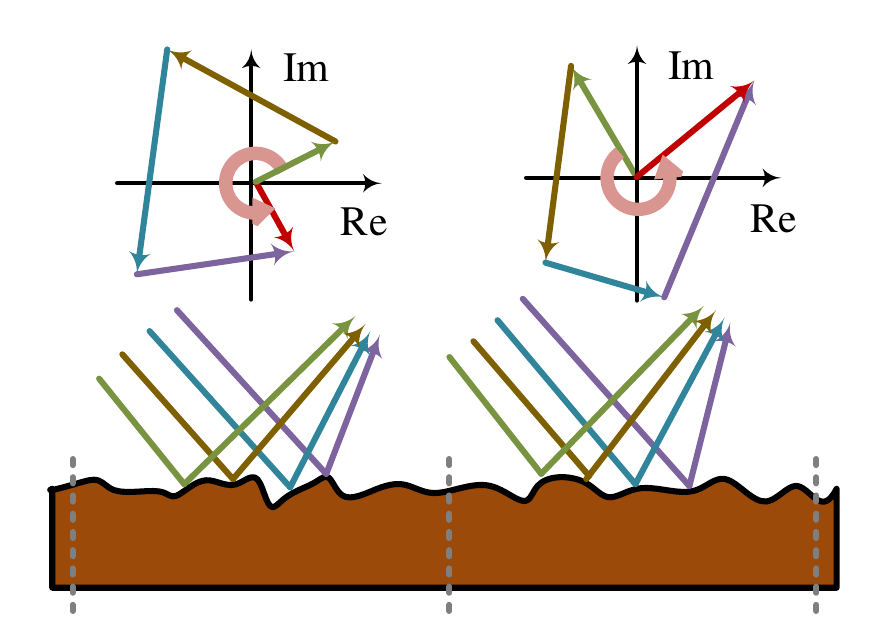}\\
	\caption{Scattering model explaining  speckle effect. Due to the coherent
		sum of many elemental scatterers within a resolution cell, the phase of the pixels are i.i.d variables uniformly distributed in $[-\pi,\pi]$ when the radar wavelength is comparable to the surface roughness \cite{Moreira2013A}. This effect is inherent in coherent imaging systems.}\label{speckle}
\end{figure}
\begin{figure}
	\centering
	\subfigure{
		\includegraphics[width=3.7cm,height=3.5cm]{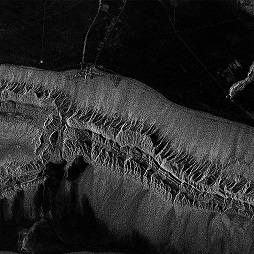}}
	\subfigure{
		\includegraphics[width=3.7cm,height=3.5cm]{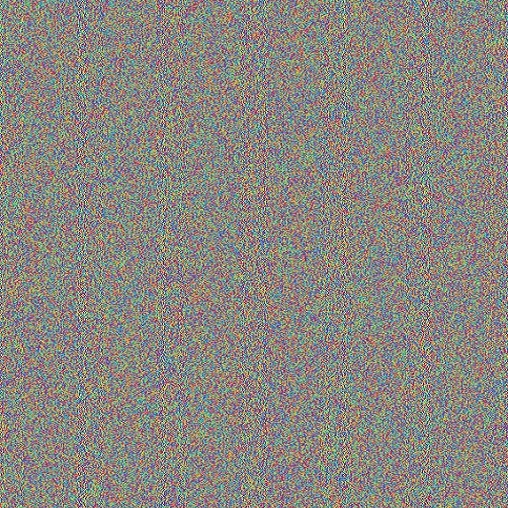}}
	\subfigure{
		\includegraphics[width=0.6cm,height=3.5cm]{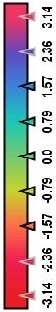}}\\
	\caption{Amplitude and phase of a SAR image. The random phase leads to a random sensing matrix. Data provided by European Space Agency].
	}\label{ap}
\end{figure}

The concentration of measure (CoM) inequality \cite{Ledoux2001The} is one of the leading techniques for RIP analyses of random matrices. It quantifies how well a random matrix preserve the norm of a high-dimensional sparse signal when mapped into a low-dimensional space. We begin by deriving the CoM inequality of ${\mathbf{A}_m}$, as presented in Lemma 1.

\vspace{0.4em}
\emph{Lemma 1:} Let $\mathbf{{x}}$ be a $k$-sparse vector in ${{\mathbb{C}}^{K}}$, then for $0<\delta <1$, we have
\begin{equation}\label{}
	\begin{split}
		\mathbb{P}\bigg\{\Big |\norm{\mathbf{A}_m\mathbf{x}}_2-\norm{\bf x}_2\Big|\ge{\delta\norm{\bf x}_2}\bigg\}
		\le4\exp \left( -\frac{{{\alpha\beta\delta^2 }}}{16{\norm{\mathbf{W}}_{k,\infty}^2}} \right),
	\end{split}
\end{equation}
where $K$ is the number of columns of $\mathbf{A}_m$, $\norm{\mathbf{W}}_{k,\infty}$ is a mixed norm of $\mathbf{W}$ defined as
\begin{equation}\label{}
	\norm{\mathbf{W}}_{k,\infty}= \sup\limits_{\norm{\bf x}_2=1,|\mathrm{supp}(\mathbf{x})|\le k} {{\norm{\bf Wx} }_{\infty}}.
\end{equation}

\vspace{0.4em}
\emph{Proof:} See Appendix A.
\vspace{0.4em}

Lemma 1 points out that the probabilistic tail {bound} of the CoM decays exponentially. It means that the sensing matrix $\mathbf{{A}}_m$ has a  distance-preserving property, which is the essence of RIP. Using covering number estimate \cite{Foucart2013A} in combination with Lemma 1, we derive the following RIP condition.

\vspace{0.4em}
\begin{thm}The restricted isometry constant ${{\delta }_{k}}$ of the sensing matrix $\mathbf{A}_m$ satisfies ${{\delta }_{k}}<\delta $ for some constant $0<\delta <1$ with probability exceeding $1-\eta $ provided that
	\begin{equation}\label{RIP}
\alpha\beta\ge \frac{C\norm{\mathbf{W}}_{k,\infty}^2}{{\delta}^{2}}\Big(k\,\mathrm{log}(eK/k)+k\,\mathrm{log}\left(36/\delta\right)+\mathrm{log}\left({4 }/{\eta}\right)\Big),
	\end{equation}
\end{thm}
where $C=576$ is a constant.

\vspace{0.4em}
\emph{Proof:} See Appendix B.
\vspace{0.4em}

According to Theorem 1, if the resolution ratios $\alpha$ and $\beta$ are properly designed, then with high probability, the sensing matrix $\mathbf{A}_m$ satisfies RIP.
The basis $\mathbf{W}$ also affects RIP. For the DCT basis, it can be given that $\|{\mathbf{W}}\|_{k,\infty}^2={{2k}/{K}}$. Thus in this case, the {sensing} matrix satisfies RIP with $\alpha\beta=\mathcal{O}\left(k^2\mathrm{log}(K/k)/K\right)$.
Namely, the total resolution ratio $\alpha\beta$ is mainly limited by the ratio between the square of interferogram sparsity and the master image size.
If interferograms are sparse enough, $\alpha\beta$ {can be much smaller} than $1$.
Then the resolution of the slave image can be much coarser than that of the master image.

Although inferior to the optimal RIP $\mathcal{O}\left(k\mathrm{log}(K/k)/K\right)$ of Gaussian matrix \cite{candes2008introduction}, the quadratic RIP in (\ref{RIP}) still shows that $\mathbf{A}_m$ is an effective sensing matrix.
According to our experience on the RIP analysis of CS radar \cite{2014Xi&Chen,2017Chen&Xi,Chen2019CSPD}, {the quadratic factor $k^2$ is possibly  overestimated by the adopted proof technique. How to theoretically tighten this RIP will be considered in our future works.}

\subsection{Connections to Random Convolution }
The measurement model (\ref{zmu}) of NCB-InSAR is similar to the CS models of \emph{random convolution} and \emph{random demodulation}  \cite{Romberg2008Compressive,Tropp}. These CS models measure sparse signals at low rate by {convoluting it} with a random sequence followed by subsampling. The subsamples can be drawn at either random locations or consecutive ones, leading to different variants of measurement matrices \cite{Romberg2008Compressive,rauhut2009circulant,Tropp}. This measurement process can be described by a measurement matrix $\mathbf{H}_{rc}=\mathbf{\Omega}\mathbf{C}$, where $\mathbf{C}$ is a random convolution (or circulant) matrix and $\mathbf{\Omega}$ is a row submatrix matrix of an identity matrix.
By diagonalizing the circulant matrix $\mathbf{C}$ via DFT, the random convolution matrix can be rewritten as
\begin{equation}\label{rd}
{\mathbf{H}_{rc}=\mathbf{\Omega}\mathbf{F}^* \mathrm{diag}(\bm{\theta})\mathbf{F}}
\end{equation}
where $\bm{\theta}$ is a random sequence consisting of complex numbers with random phases \cite{Romberg2008Compressive}. Actually, NCB-InSAR adopts a similar measurement model, as stated in Proposition 1.

\vspace{0.4em}
\emph{Proposition 1:} Denote $\mathbf{\Omega}_{2D}=\mathbf{\Omega}_{\beta}\otimes\mathbf{\Omega}_{\alpha}$ and $\mathbf{F}_{2D}={\mathbf{F}}_{L}\otimes\mathbf{F}_{N}$. The measurement model (\ref{zmu}) of NCB-InSAR can be reformulated as
\begin{equation}\label{rd0}
	\hat{\mathbf{y}}_s=\hat{\mathbf{H}}_m\hat{\mathbf{u}}
\end{equation}
where
\begin{equation}\label{rd2} \hat{\mathbf{H}}_m=\frac{1}{\sqrt{\alpha\beta}}\mathbf{\Omega}_{2D}\mathbf{F}_{2D}\mathrm{diag}(\mathbf{\Theta}_m)\mathbf{F}_{2D}^*
\end{equation}
is a 2-D random convolution matrix, and $\hat{\mathbf{y}}_s=(\mathbf{F}_{J}\otimes\mathbf{F}_{I})\mathbf{y}_s$,  $\hat{\mathbf{u}}=\mathbf{F}_{2D}{\mathbf{u}}$.
\vspace{0.4em}

\emph{Proof:}
Firstly, we decompose the Kronecker product bracketed in (\ref{mesmtx}) as
\begin{equation}\label{}
	\begin{split}
		&\mathbf{F}_{J}^* \mathbf{\Omega}_{\beta}{\mathbf{F}}_{L}\otimes\mathbf{F}_{I}^*\mathbf{\Omega}_{\alpha}\mathbf{F}_{N}\\
		& =(\mathbf{F}_{J}^*\otimes\mathbf{F}_{I}^*)
		(\mathbf{\Omega}_{\beta}\otimes\mathbf{\Omega}_{\alpha})({\mathbf{F}}_{L}\otimes\mathbf{F}_{N})\\
		&=(\mathbf{F}_{J}^*\otimes\mathbf{F}_{I}^*)\mathbf{\Omega}_{2D}\mathbf{F}_{2D}.
	\end{split}
\end{equation}
Next, we multiply (\ref{zmu}) by $(\mathbf{F}_{J}\otimes\mathbf{F}_{I})$ and denote $ \hat{\mathbf{u}}=\mathbf{F}_{2D}\mathbf{u}$.  It then results in {the} measurement model  $\hat{\mathbf{y}}_s=\hat{\mathbf{H}}_m\hat{\mathbf{u}}$ with
$\hat{\mathbf{H}}_m=\frac{1}{\sqrt{\alpha\beta}}\mathbf{\Omega}_{2D}\mathbf{F}_{2D}\mathrm{diag}(\mathbf{\Theta}_m)\mathbf{F}_{2D}^*$.
\vspace{0.4em}

Observing (\ref{rd}) and (\ref{rd2}), we find that $\hat{\mathbf{H}}_m$ has a similar structure to the random convolution matrix $\mathbf{H}_{rc}$. The difference is that the random convolution and subsampling of $\hat{\mathbf{H}}_m$ is performed in two-dimensional frequency domain while (\ref{rd}) is in one-dimensional time domain.
Since random convolution matrices have been proved to satisfy RIP \cite{Romberg2008Compressive,rauhut2009circulant}, it indirectly confirms the effectiveness of our CS model.

In summary, both Theorem 1 and Proposition 1 show that the sensing matrix of NCB-InSAR has an effective structure to satisfy RIP. So, from the viewpoint of CS, the HR interferogram is stably embedded into the NCB-InSAR image pair, and thus {it is feasible to be recovered correctly and robustly via $\ell_1$ minimization.}

\section{NIL1M Algorithm}
This section is to {design} an efficient interferogram formation algorithm for NCB-InSAR. Actually, there exist many standard solvers available for the problems (\ref{socp}) and (\ref{bpdn}), such as TFOCS \cite{Becker2011} and $\ell_1$-MAGIC \cite{Candes2005l1}. However, these solvers are developed for general problems, not suitable for our problem considering {that our problem size is typically large due to} large image sizes. Therefore, we design the NIL1M algorithm for fast interferogram formation.

\subsection{Problem Reformulation}
We focus on solving the $\ell_1$ minimization problem (\ref{bpdn}), considering that it is easier to handle compared to the quadratically constrained problem (\ref{socp}).
To exploit the DFT structures of the sensing matrix, we rewrite (\ref{bpdn}) in a matrix form as
\begin{equation}\label{bpdn2}
	\min\limits_{\mathbf{U}}^{}   \left\|{\mathbf{Y}_s}-\mathcal{H}_m\mathbf{U}\right\|_F^2+\lambda\norm{\mathcal{W}\mathbf{U}}_1 ,
\end{equation}
where $\mathcal{H}_m$ denotes the measurement operator,
\begin{equation}\label{}
	\mathcal{H}_m\mathbf{U}=\frac{1}{\sqrt{\alpha\beta}}\mathbf{F}_{I}^*\mathbf{\Omega}_{\alpha}\mathbf{F}_{N} (\mathbf{\Theta}_m\circ\mathbf{U}) \mathbf{F}_{L}^T\mathbf{\Omega}_{\beta}^T\overline{\mathbf{F}}_{J},
\end{equation}
and $\mathcal{W}$ is a sparse transform, e.g. wavelet transform or DCT. It is notable that, this kind of optimization problems is related to image {deblurring/deconvolution} approaches \cite{Figueiredo2003An,beck2009fastcof}.

First-order methods are commonly used to solve this kind of large-scale problems, such as the family of iterative
shrinkage/thresholding algorithms \cite{KHORAMIAN2012109,beck2009fast}. In these algorithms, each iteration involves the measurement operator $\mathcal{H}_m$ and its adjoint, contributing to main computational burden. So, it is helpful to simplify the operator $\mathcal{H}_m$ beforehand to reduce the computational cost.
To this end, we apply a simple pre-processing to the slave image $\mathbf{Y}_s$, i.e., a 2-D DFT operation,
\begin{equation}\label{prep}
	\hat{\mathbf{Y}}_s=\mathbf{F}_I \mathbf{Y}_s \mathbf{F}^T_J.
\end{equation}
Since all DFT matrices are orthonormal, we can replace $\norm{{\mathbf{Y}_s}-\mathcal{H}_m\mathbf{U}}^2_F$ in (\ref{bpdn2}) by
\begin{equation}\label{fU}
	f(\mathbf{U}) = \norml{{\hat{\mathbf{Y}}_s}-\hat{\mathcal{H}}_m\mathbf{U}}_F^2,
\end{equation}
where
\begin{equation}\label{}
	\hat{\mathcal{H}}_m\mathbf{U}=\frac{1}{\sqrt{\alpha\beta}}\mathbf{\Omega}_{\alpha}\mathbf{F}_{N} (\mathbf{\Theta}_m\circ\mathbf{U}) \mathbf{F}_{L}^T\mathbf{\Omega}_{\beta}^T.
\end{equation}
This replacement saves the DFTs involving  $\mathbf{F}_{I}$ and $\mathbf{F}_{J}$.
Now, we have an equivalent form of the problem (\ref{bpdn2}),
\begin{equation}\label{bpdn3}
	\min\limits_{\mathbf{U}}^{}   f(\mathbf{U})+\lambda\norm{\mathcal{W}\mathbf{U}}_1.
\end{equation}

\subsection{Algorithm Design}
With the equivalent formulation (\ref{bpdn3}), we design a first-order interferogram formation algorithm under the framework of the \emph{fast iterative shrinkage-thresholding algorithm} (FISTA) \cite{beck2009fast} in virtue of its simplicity and efficiency.

Using quadratic approximation of $f(\mathbf{U})$, each iteration of FISTA for (\ref{bpdn3}) is to solve the following subproblem \cite{beck2009fast},
\begin{equation}\label{subproblem0}
	\begin{split}
		\mathbf{U}_i=\mathop{\mathrm{argmin}}\limits_{\mathbf{U}}\,\bigg\{&f(\mathbf{V}_i)+\langle\mathbf{U}-\mathbf{V}_i,\nabla f(\mathbf{V}_i)\rangle \\
		& +\frac{L_f}{2}\norm{\mathbf{U}-\mathbf{V}_i}_F^2+\lambda\norm{\mathcal{W}\mathbf{U}}_1\bigg\},
	\end{split}
\end{equation}
where $L_f$ is the Lipschitz constant of the gradient $\nabla f(\mathbf{U})$ of $f(\mathbf{U})$, and $\mathbf{V}_i$ is a linear combination of previous two iterative points $\{\mathbf{U}_{i-1},\mathbf{U}_{i-2}\}$. {In Appendix C, we derive that $L_f=2/\alpha\beta$.} $\mathbf{V}_i$ will be given explicitly in later discussions.
After completing the square and ignoring constant terms, the subproblem (\ref{subproblem0}) becomes
\begin{equation}\label{subproblem}
	\begin{split}
		\mathbf{U}_i=\mathop{\mathrm{argmin}}\limits_{\mathbf{U}}\,\bigg\{&
		\frac{1}{2}\left\|\mathbf{U}-\left(\mathbf{V}_i-\frac{1}{L_f}\nabla f(\mathbf{V}_i)\right)\right\|_F^2\\
		&+\frac{\lambda}{L_f}\norm{\mathcal{W}\mathbf{U}}_1\bigg\}.
	\end{split}
\end{equation}
Using the soft-thresholding operator
\begin{equation}\label{}
	\mathrm{soft}(\mathbf{V},\kappa)_{n,l}=\mathrm{sign}(v_{n,l})\cdot \max\big\{|v_{n,l}|-\kappa,0\big\},
\end{equation}
where $v_{n,l}$ denotes the $[n,l]$-th entry of $\mathbf{V}$, and $\kappa$ is the soft-threshold, the subproblem (\ref{subproblem}) has a closed-form solution as
\begin{equation}\label{}
	\mathbf{U}_i=\mathcal{W}^*\mathrm{soft}\left(\mathcal{W}\left (\mathbf{V}_i-\mathbf{D}_i\right),\frac{\lambda}{L_f}\right),
\end{equation}
where $\mathbf{D}_i=\nabla f(\mathbf{V}_i)/{L_f}$. According to the definition of $f(\mathbf{U})$ in (\ref{fU}), $\mathbf{D}_i$ can be given explicitly as
\begin{equation}\label{}
	\begin{split} \mathbf{D}_i&=-\frac{2}{L_f}\hat{\mathcal{H}}_m^*\left({\hat{\mathbf{Y}}_s}-\hat{\mathcal{H}}_m\mathbf{V}_i\right)\\
		&=\frac{2}{\sqrt{\alpha\beta}L_f}{\overline{\mathbf{\Theta}}_m} \circ   \mathbf{F}_{N}^*\mathbf{\Omega}_{\alpha}^T\mathbf{R}_i\mathbf{\Omega}_{\beta}\overline{\mathbf{F}}_{L},
	\end{split}
\end{equation}
where
\begin{equation}\label{}
	\mathbf{R}_i =	{\hat{\mathbf{Y}}_s}-\frac{1}{\sqrt{\alpha\beta}}\mathbf{\Omega}_{\alpha}\mathbf{F}_{N} (\mathbf{\Theta}_m\circ\mathbf{V}_i) \mathbf{F}_{L}^T\mathbf{\Omega}_{\beta}^T.
\end{equation}
Intuitively, the variable $\mathbf{R}_i$ is the spectral residual of the LR {slave} image in $i$-th iteration, and  $\mathbf{D}_i$ is a scaled projection of the residual in high-dimension interferogram space.

\begin{algorithm}[t]
	\caption{NIL1M Algorithm.}
	\label{alg:Framwork}
	\begin{algorithmic}[1]
		\STATE initial: take $L_f=2/\alpha\beta$, $\kappa=\lambda/L_f$, number of iterations $N_{iter}$, and set $\mathbf{U}_0=\mathbf{V}_1=\mathbf{0}$, $t_0=1$, $i=1$;
		\STATE $\hat{\mathbf{Y}}_s=\mathbf{F}_I \mathbf{Y}_s \mathbf{F}^T_J$;
		\REPEAT
		\STATE $\mathbf{R}_i =	{\hat{\mathbf{Y}}_s}-\frac{1}{\sqrt{\alpha\beta}}\mathbf{\Omega}_{\alpha}\mathbf{F}_{N} (\mathbf{\Theta}_m\circ\mathbf{V}_i) \mathbf{F}_{L}^T\mathbf{\Omega}_{\beta}^T $;
		\STATE $\mathbf{D}_i=-\frac{2}{\sqrt{\alpha\beta}L_f}{\overline{\mathbf{\Theta}}_m} \circ   \mathbf{F}_{N}^*\mathbf{\Omega}_{\alpha}^T\mathbf{R}_i\mathbf{\Omega}_{\beta}\overline{\mathbf{F}}_{L}$;
		\STATE $\mathbf{U}_i=\mathcal{W}^*\mathrm{soft}\left( \mathcal{W}\left(\mathbf{V}_i-\mathbf{D}_i\right),\kappa\right)$;
		\STATE $t_i={\frac{1}{2}+\frac{1}{2}\sqrt{1+4t_{i-1}^2}}$;\\
		\STATE $\mathbf{V}_{i+1}=\mathbf{U}_i+\frac{t_{i-1}-1}{t_i}(\mathbf{U}_i-\mathbf{U}_{i-1})$
		\STATE        $i=i+1$;
		\UNTIL $i> N_{iter}$;
       \STATE {Take the angle of $\mathbf{U}_i$ to obtain $\bm\phi_{topo}$.}
	\end{algorithmic}
\end{algorithm}

Based on the discussions above, we formulate the NIL1M algorithm in Algorithm 1.
In the algorithm, the computational complexity mainly {lies in} the involved DFTs and sparse transforms. By exploiting the FFT technique and fast implementation of wavelet transforms, the algorithm has an $\mathcal{O}\left(K\mathrm{log}(K)\right)$ complexity per iteration. If we solve the original problem (\ref{bpdn}) directly, the computational complexity should be $\mathcal{O}(K^2)$ per iteration for common first-order solvers.
Therefore, our algorithm is more efficient to solve (\ref{bpdn}) with large scales.

\subsection{Parameter Selection}
{Regarding the regularization parameter $\lambda$, it is related to the Lagrange multiplier of the constraint in (\ref{socp}). The optimal choice of $\lambda$ should provide a solution that coincides with the solution of (\ref{socp}). However, such choice cannot be known a prior \cite{van2008probing}.  Chen et al. suggest a strategy for choosing the regularization parameter, i.e., $ \lambda=\sigma\sqrt{2\,\mathrm{log} K}$,
where $\sigma$ is the noise level \cite{chen2001atomic}. This choice has some optimal properties based on assumptions of the orthogonality of ${\bf A}_m$ and the known noise level. For our problem, ${\bf A}_m$ is not orthogonal. As a relaxation, we suggest to obtain suitable $\lambda$ using a strategy similar to $\sigma\sqrt{2\,\mathrm{log} K}$.  Specifically, defining a positive hyperparameter $\gamma$ and
\begin{equation}\label{sigma}
  \sigma_{\gamma}=\sqrt{\frac{1}{\gamma}\cdot\frac{\norm{\mathbf{y}_s}^2_2}{\mathrm{length}(\mathbf{y}_s)}}
  \end{equation}
 where $\mathrm{length}(\cdot)$ denotes the sample length of a vector, we set
 \begin{equation}\label{lambda}
   \lambda=\sigma_{\gamma}\sqrt{2\,\mathrm{log} K}.
\end{equation}
If $\gamma$ is set as  signal-to-noise ratio (SNR), $\sigma_{\gamma}$ will also lead to the strategy $\lambda=\sigma\sqrt{2\,\mathrm{log} K}$. For our problem, we suggest selecting $\gamma$ around $1$, for example, $\gamma=1, 0.5,0.25$. According to our experience, it is easy to obtain relatively suitable $\lambda$ by several trials of the  hyperparameter $\gamma$. }

\begin{figure*}[t]
	\centering
{\includegraphics[scale=0.3]{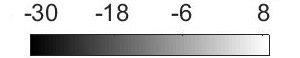}\hspace{6.2em}
\includegraphics[scale=0.3]{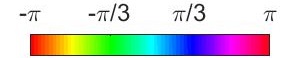}\hspace{9.3em}
\includegraphics[scale=0.3]{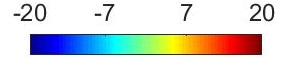}\hspace{0.5em}
\includegraphics[height=0.2cm,width=0.9cm]{colormap//padding.pdf}}\\
\subfigure[Image amplitude and phase.]{\includegraphics[width=2.13cm]{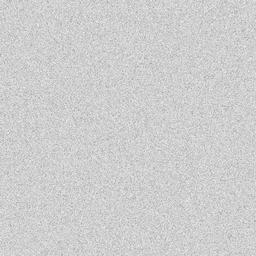}
\includegraphics[width=2.13cm]{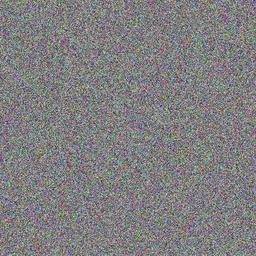}}\hspace{0.12em}
\subfigure[Example 1.]{
\includegraphics[width=2.13cm]{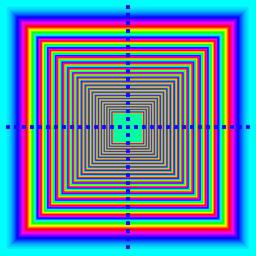}}\hspace{0.12em}
\subfigure[Example 2.]{
\includegraphics[width=2.13cm]{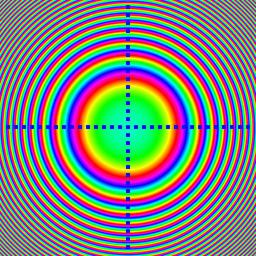}}\hspace{0.12em}
\subfigure[DCT of topographic phases.]{\includegraphics[width=2.13cm]{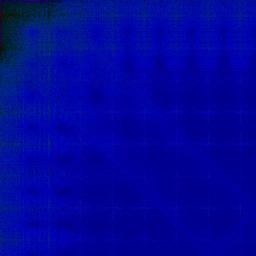}
\includegraphics[width=2.13cm]{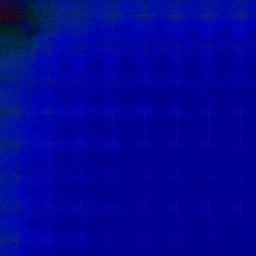}}
	\caption{Simulated examples with $1024\times 1024$ pixels. (a) simulated image amplitude and random phase; (b) the first simulated topography phase; (c) the second simulated topography phase; (c) the DCT coefficients of the simulated complex exponential topographic phases.  The range is horizontal, and the azimuth is vertical.}\label{sim_ref}
\end{figure*}

\begin{figure*}
  \centering
\includegraphics[scale=0.3]{colormap//phase_colormap.jpg}\\
  \subfigure[Recovery of example 1.]{\begin{overpic}[width=2.13cm]{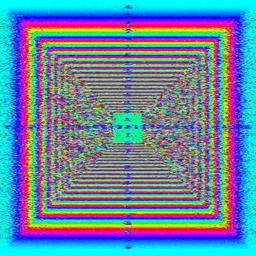}
        \put(65,50){\setlength\fboxrule{1pt}\setlength{\fboxsep}{3.2pt}\color{green}\fbox{}}
         \put(42.4,28){\setlength\fboxrule{1pt}\setlength{\fboxsep}{3.6pt}\color{yellow}\fbox{}}
	\end{overpic}
 \begin{overpic}[width=2.13cm]{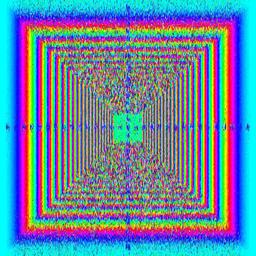}
      \put(65,50){\setlength\fboxrule{1pt}\setlength{\fboxsep}{3.2pt}\color{green}\fbox{}}
          \put(42.4,28){\setlength\fboxrule{1pt}\setlength{\fboxsep}{3.2pt}\color{yellow}\fbox{}}
	\end{overpic}
   \begin{overpic}[width=2.13cm]{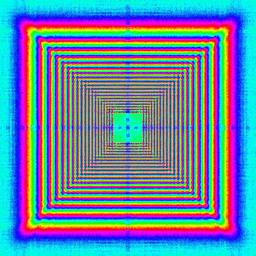}
      \put(65,50){\setlength\fboxrule{1pt}\setlength{\fboxsep}{3.2pt}\color{green}\fbox{}}
          \put(42.4,28){\setlength\fboxrule{1pt}\setlength{\fboxsep}{3.2pt}\color{yellow}\fbox{}}
	\end{overpic}
   \begin{overpic}[width=2.13cm]{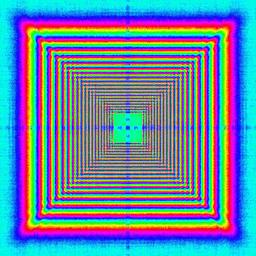}
      \put(65,50){\setlength\fboxrule{1pt}\setlength{\fboxsep}{3.2pt}\color{green}\fbox{}}
          \put(42.4,28){\setlength\fboxrule{1pt}\setlength{\fboxsep}{3.2pt}\color{yellow}\fbox{}}
	\end{overpic}}\hspace{0.25em}
\subfigure[Recovery of example 2.]{\begin{overpic}[width=2.13cm]{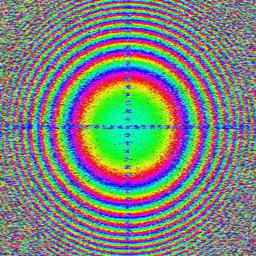}
    \put(66,50){\setlength\fboxrule{1pt}\setlength{\fboxsep}{3.2pt}\color{green}\fbox{}}
       \put(86,7){\setlength\fboxrule{1pt}\setlength{\fboxsep}{3.2pt}\color{yellow}\fbox{}}
	\end{overpic}
 \begin{overpic}[width=2.13cm]{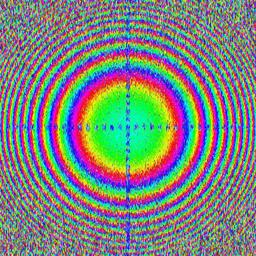}
        \put(66,50){\setlength\fboxrule{1pt}\setlength{\fboxsep}{3.2pt}\color{green}\fbox{}}
         \put(86,7){\setlength\fboxrule{1pt}\setlength{\fboxsep}{3.2pt}\color{yellow}\fbox{}}
	\end{overpic}
\begin{overpic}[width=2.13cm]{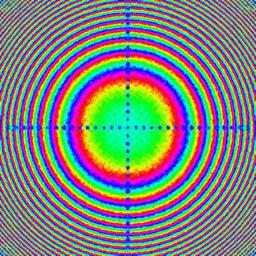}
          \put(66,50){\setlength\fboxrule{1pt}\setlength{\fboxsep}{3.2pt}\color{green}\fbox{}}
           \put(86,7){\setlength\fboxrule{1pt}\setlength{\fboxsep}{3.2pt}\color{yellow}\fbox{}}
	\end{overpic}
    \begin{overpic}[width=2.13cm]{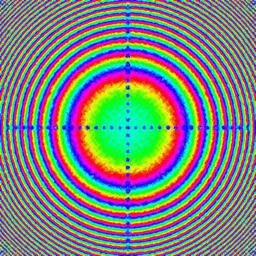}
     \put(66,50){\setlength\fboxrule{1pt}\setlength{\fboxsep}{3.2pt}\color{green}\fbox{}}
     \put(86,7){\setlength\fboxrule{1pt}\setlength{\fboxsep}{3.2pt}\color{yellow}\fbox{}}
	\end{overpic}}\\
         \subfigure[Yellow-framed areas in (a).]{\includegraphics[width=2.13cm]{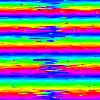}
    \includegraphics[width=2.13cm]{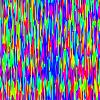}
      \includegraphics[width=2.13cm]{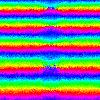}
    \includegraphics[width=2.13cm]{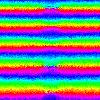}}\hspace{0.25em}
           \subfigure[Yellow-framed areas in (b).]{\includegraphics[width=2.13cm]{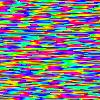}
    \includegraphics[width=2.13cm]{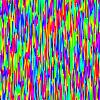}
      \includegraphics[width=2.13cm]{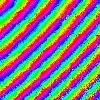}
    \includegraphics[width=2.13cm]{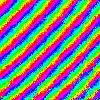}}\\
         \subfigure[Green-framed areas in (a).]{\includegraphics[width=2.13cm]{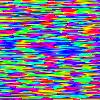}
    \includegraphics[width=2.13cm]{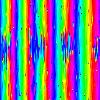}
      \includegraphics[width=2.13cm]{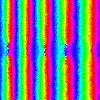}
    \includegraphics[width=2.13cm]{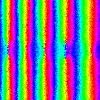}}\hspace{0.25em}
           \subfigure[Green-framed areas in (b).]{\includegraphics[width=2.13cm]{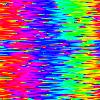}
    \includegraphics[width=2.13cm]{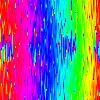}
      \includegraphics[width=2.13cm]{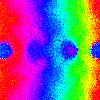}
    \includegraphics[width=2.13cm]{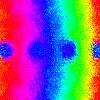}}\\
  \caption{Simulations in ideal noise-free setting. In each subfigure, the first and second images are processed by CB-InSAR with resolution ratios $1/16\times 1$ and $1\times 1/16$, respectively; the third and fourth images are processed by NCB-InSAR with resolution ratios $1/16\times 1$ and $1\times 1/16$, respectively. {Notice that NCB-InSAR can better recover the homogeneous fringes, but does not preserve the small phase patches ($16\times 16$ pixels) as shown in (c) and (e).}
 }\label{sim_noisefree}
\end{figure*}

\begin{figure*}
  \centering
\includegraphics[scale=0.3]{colormap//phase_colormap.jpg}\\
  \subfigure[Recovery of example 1.]{\begin{overpic}[width=2.13cm]{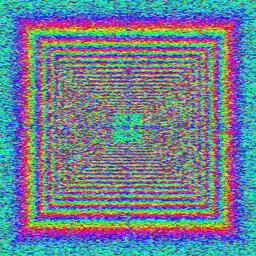}
              \put(65,50){\setlength\fboxrule{1pt}\setlength{\fboxsep}{3.2pt}\color{green}\fbox{}}
          \put(42.4,28){\setlength\fboxrule{1pt}\setlength{\fboxsep}{3.2pt}\color{yellow}\fbox{}}
	\end{overpic}
 \begin{overpic}[width=2.13cm]{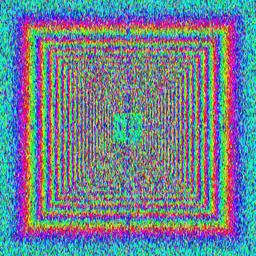}
        \put(65,50){\setlength\fboxrule{1pt}\setlength{\fboxsep}{3.2pt}\color{green}\fbox{}}
          \put(42.4,28){\setlength\fboxrule{1pt}\setlength{\fboxsep}{3.2pt}\color{yellow}\fbox{}}
	\end{overpic}
   \begin{overpic}[width=2.13cm]{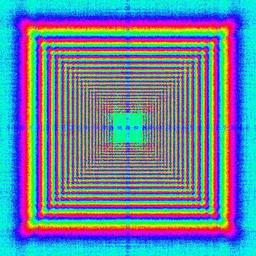}
            \put(65,50){\setlength\fboxrule{1pt}\setlength{\fboxsep}{3.2pt}\color{green}\fbox{}}
          \put(42.4,28){\setlength\fboxrule{1pt}\setlength{\fboxsep}{3.2pt}\color{yellow}\fbox{}}
	\end{overpic}
   \begin{overpic}[width=2.13cm]{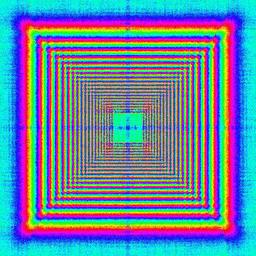}
            \put(65,50){\setlength\fboxrule{1pt}\setlength{\fboxsep}{3.2pt}\color{green}\fbox{}}
          \put(42.4,28){\setlength\fboxrule{1pt}\setlength{\fboxsep}{3.2pt}\color{yellow}\fbox{}}
	\end{overpic}}\hspace{0.25em}
\subfigure[Recovery of example 2.]{\begin{overpic}[width=2.13cm]{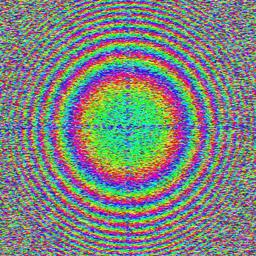}
    \put(66,50){\setlength\fboxrule{1pt}\setlength{\fboxsep}{3.2pt}\color{green}\fbox{}}
       \put(86,7){\setlength\fboxrule{1pt}\setlength{\fboxsep}{3.2pt}\color{yellow}\fbox{}}
	\end{overpic}
 \begin{overpic}[width=2.13cm]{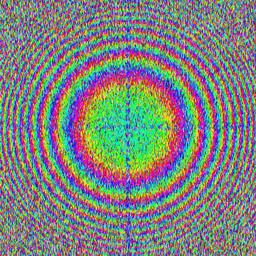}
        \put(66,50){\setlength\fboxrule{1pt}\setlength{\fboxsep}{3.2pt}\color{green}\fbox{}}
         \put(86,7){\setlength\fboxrule{1pt}\setlength{\fboxsep}{3.2pt}\color{yellow}\fbox{}}
	\end{overpic}
\begin{overpic}[width=2.13cm]{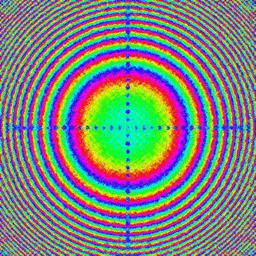}
          \put(66,50){\setlength\fboxrule{1pt}\setlength{\fboxsep}{3.2pt}\color{green}\fbox{}}
           \put(86,7){\setlength\fboxrule{1pt}\setlength{\fboxsep}{3.2pt}\color{yellow}\fbox{}}
	\end{overpic}
    \begin{overpic}[width=2.13cm]{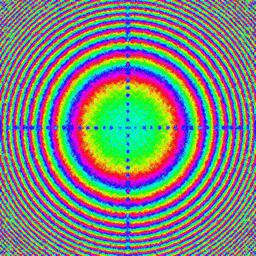}
    \put(66,50){\setlength\fboxrule{1pt}\setlength{\fboxsep}{3.2pt}\color{green}\fbox{}}
     \put(86,7){\setlength\fboxrule{1pt}\setlength{\fboxsep}{3.2pt}\color{yellow}\fbox{}}
	\end{overpic}}\\
         \subfigure[Yellow-framed areas in (a).]{\includegraphics[width=2.13cm]{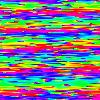}
    \includegraphics[width=2.13cm]{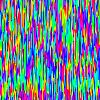}
      \includegraphics[width=2.13cm]{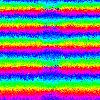}
    \includegraphics[width=2.13cm]{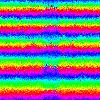}}\hspace{0.25em}
           \subfigure[Yellow-framed areas in (b).]{\includegraphics[width=2.13cm]{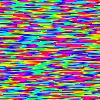}
    \includegraphics[width=2.13cm]{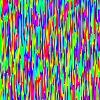}
      \includegraphics[width=2.13cm]{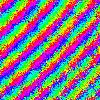}
    \includegraphics[width=2.13cm]{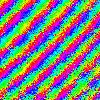}}\\
         \subfigure[Green-framed areas in (a).]{\includegraphics[width=2.13cm]{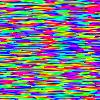}
    \includegraphics[width=2.13cm]{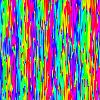}
      \includegraphics[width=2.13cm]{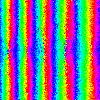}
    \includegraphics[width=2.13cm]{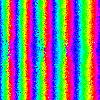}}\hspace{0.25em}
           \subfigure[Green-framed areas in (b).]{\includegraphics[width=2.13cm]{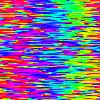}
    \includegraphics[width=2.13cm]{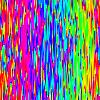}
      \includegraphics[width=2.13cm]{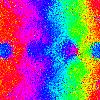}
    \includegraphics[width=2.13cm]{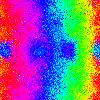}}\\
  \caption{Simulations in noisy setting, where $\phi_{noise}$ is i.i.d. random noise uniformly distributed in $[-\frac{\pi}{4},\frac{\pi}{4}]$. In each subfigure, the first and second images are processed by CB-InSAR with resolution ratios $1/16\times 1$ and $1\times 1/16$, respectively; the third and fourth images are processed by NCB-InSAR with resolution ratios $1/16\times 1$ and $1\times 1/16$, respectively.}\label{sim_noisy}
\end{figure*}

\begin{figure*}
  \centering
      \includegraphics[scale=0.3]{colormap//wav_err_colormap.jpg}\\
      \subfigure[Example 1, noise-free setting.]{\includegraphics[width=2.13cm]{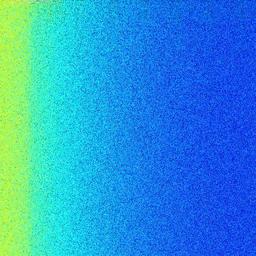}
    \includegraphics[width=2.13cm]{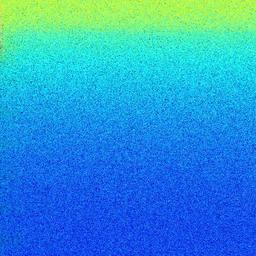}
      \includegraphics[width=2.13cm]{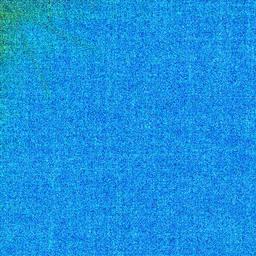}
    \includegraphics[width=2.13cm]{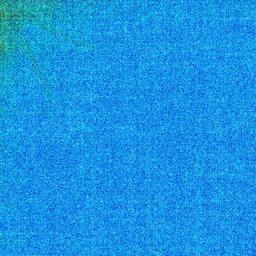}}\hspace{0.25em}
     \subfigure[Example 2, noise-free setting.]{\includegraphics[width=2.13cm]{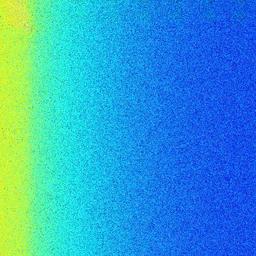}
    \includegraphics[width=2.13cm]{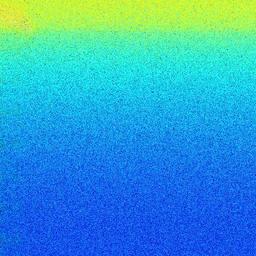}
      \includegraphics[width=2.13cm]{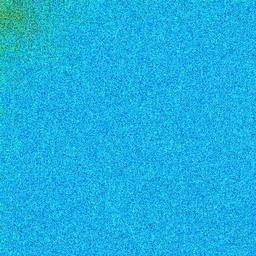}
    \includegraphics[width=2.13cm]{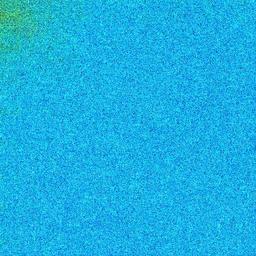}}\\
          \subfigure[Example 1, noisy setting.]{\includegraphics[width=2.13cm]{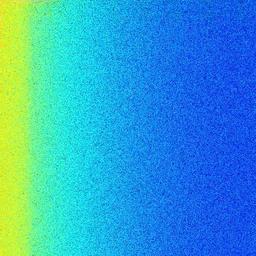}
    \includegraphics[width=2.13cm]{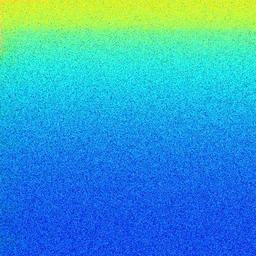}
      \includegraphics[width=2.13cm]{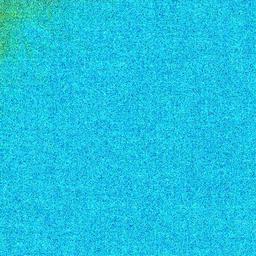}
    \includegraphics[width=2.13cm]{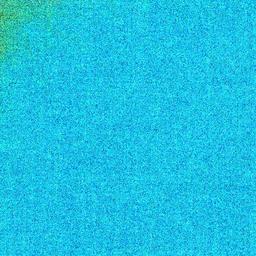}}\hspace{0.25em}
     \subfigure[Example 2, noisy setting.]{\includegraphics[width=2.13cm]{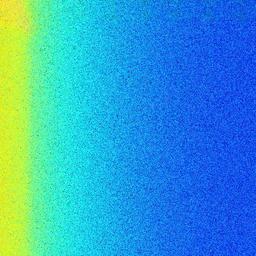}
    \includegraphics[width=2.13cm]{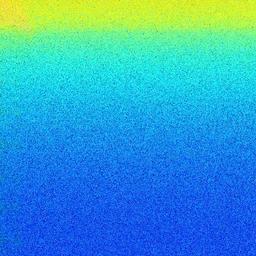}
      \includegraphics[width=2.13cm]{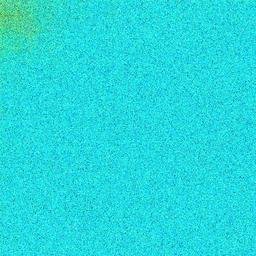}
    \includegraphics[width=2.13cm]{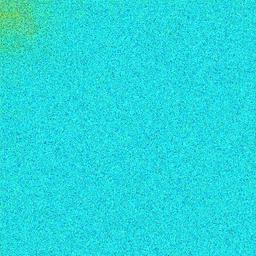}}\\
\vspace{0.5em}
   \includegraphics[width=6cm]{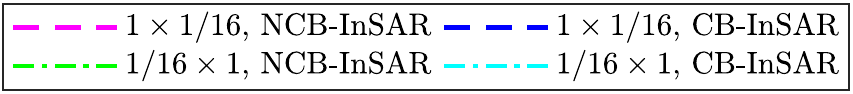}\\
    \subfigure[Example 1, noise-free setting.]{\includegraphics[width=4.4cm]{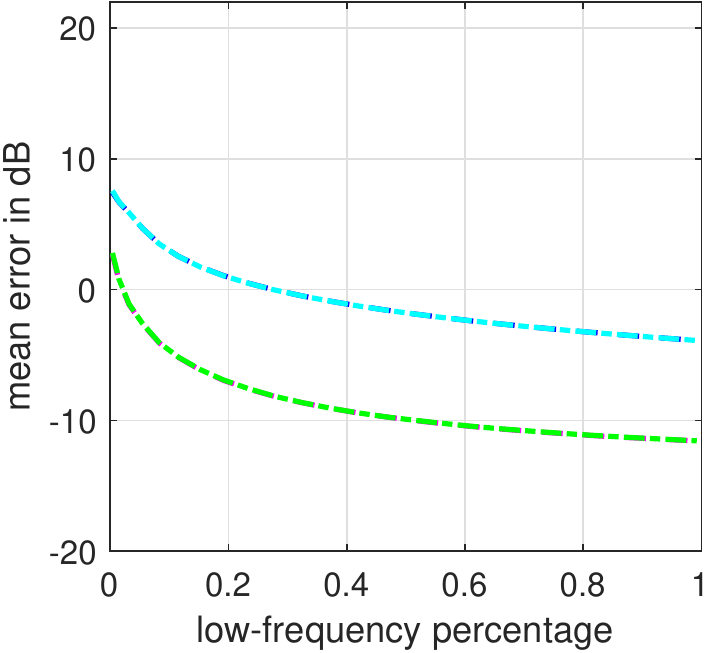}}
    \subfigure[Example 1, noisy setting.]{\includegraphics[width=4.4cm]{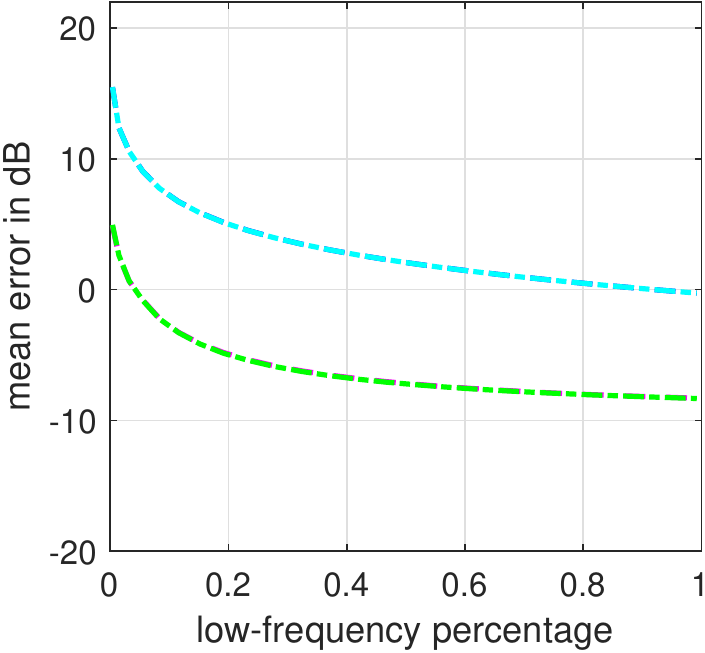}}
    \subfigure[Example 2, noise-free setting.]{\includegraphics[width=4.4cm]{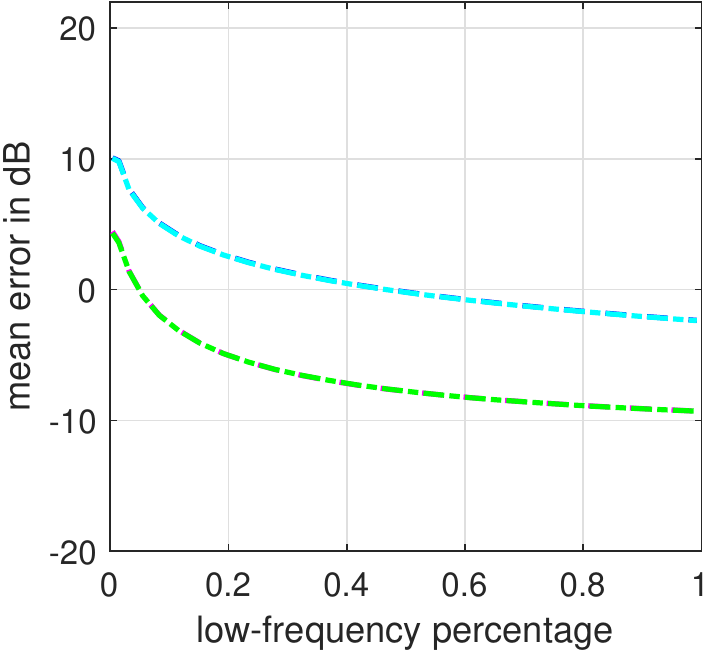}}
    \subfigure[Example 2, noisy setting.]{\includegraphics[width=4.4cm]{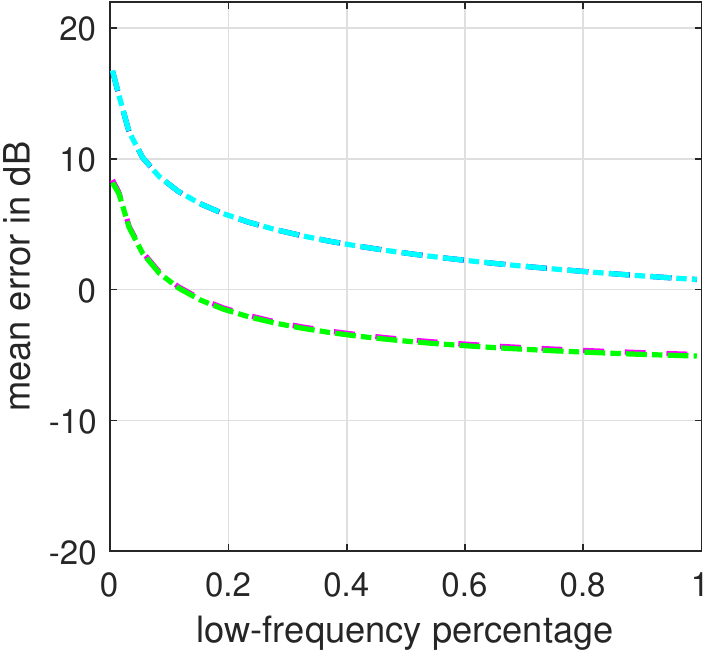}}\\
    \subfigure[Example 1, noise-free setting.]{\includegraphics[width=4.4cm]{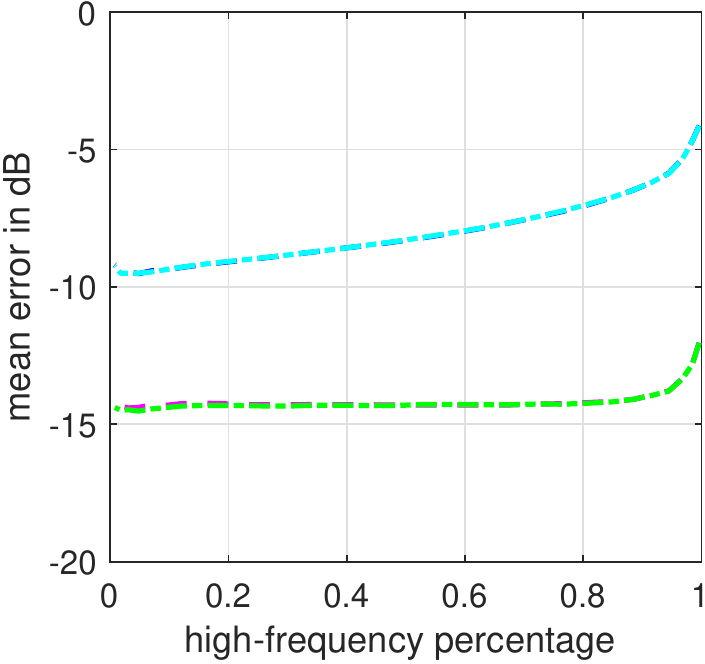}}
    \subfigure[Example 1, noisy setting.]{\includegraphics[width=4.4cm]{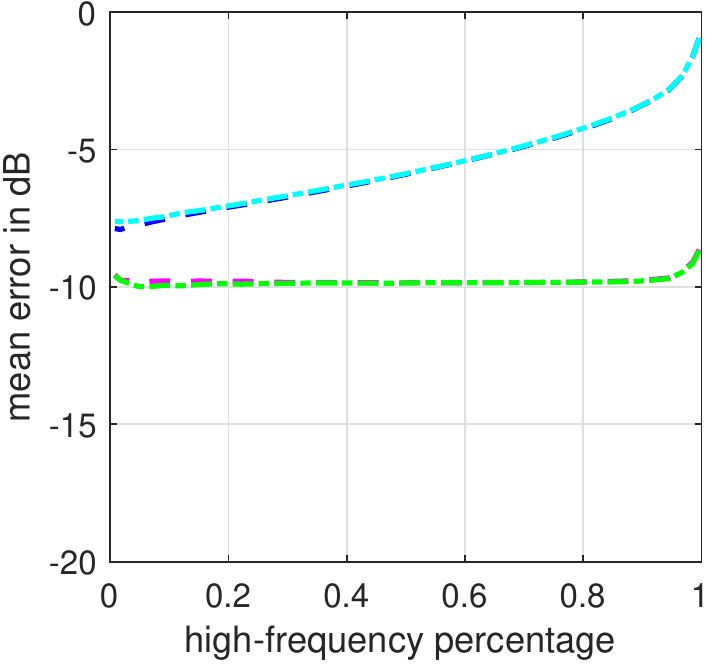}}
    \subfigure[Example 2, noise-free setting.]{\includegraphics[width=4.4cm]{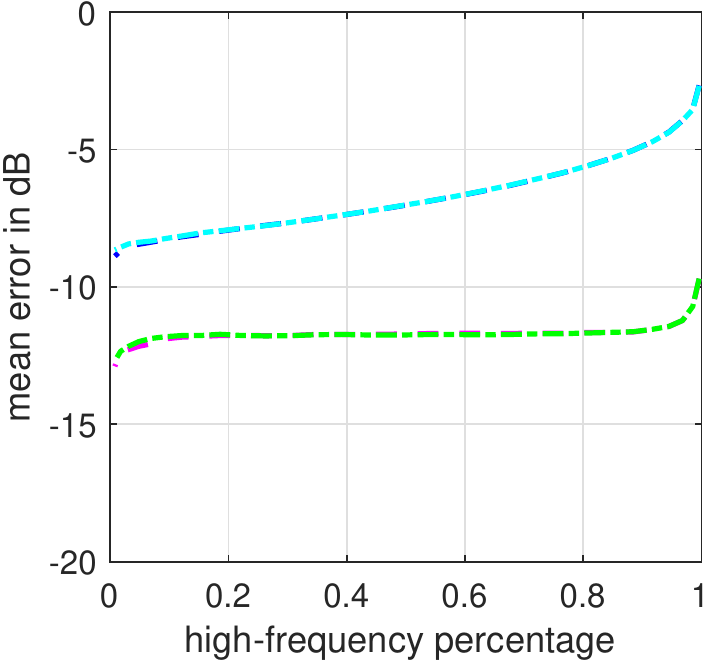}}
    \subfigure[Example 2, noisy setting.]{\includegraphics[width=4.4cm]{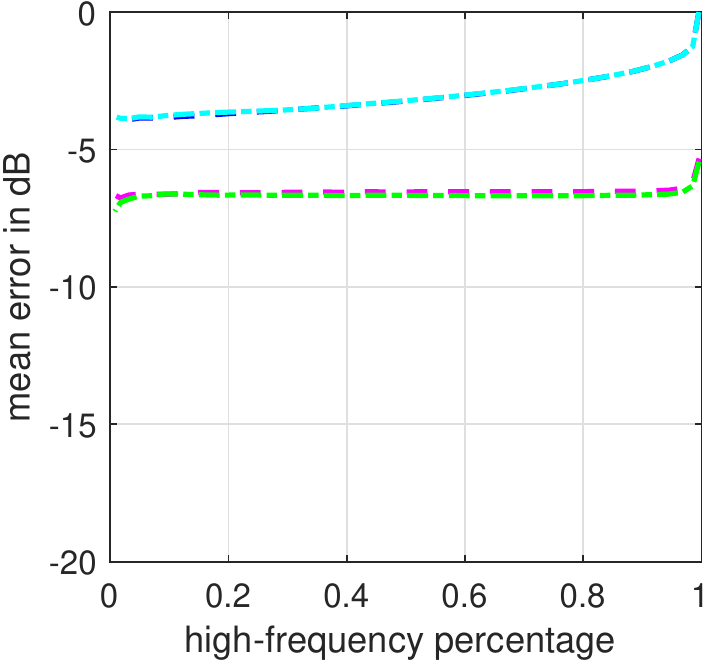}}
  \caption{ DCT coefficient errors and their distributions for tests on simulated datasets.
  {(a)-(d) show the DCT coefficient errors for Figs. \ref{sim_noisefree} (a)-(b) and Figs. \ref{sim_noisy} (a)-(b); (e)-(l) show the corresponding error distributions.} As shown above, the error curves of NCB-InSAR are below those of CB-InSAR, which means that the recovery errors of NCB-InSAR are smaller than CB results. The lower two curves are almost overlapped.}\label{sim_freq_err}
\end{figure*}

\begin{figure*}
  \centering
  {\includegraphics[scale=0.3]{colormap//phase_colormap.jpg}\hspace{5em}
\includegraphics[scale=0.3]{colormap//wav_err_colormap.jpg}
\includegraphics[height=0.1cm,width=1.3cm]{colormap//padding.pdf}\hspace{1em}}\\
  \subfigure[HR reference phases and DCT, DB-4 coefficents.]{
  \includegraphics[width=2.8cm]{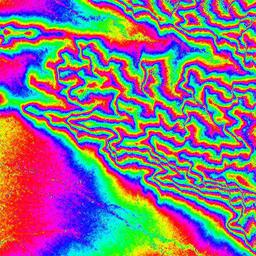}
    \includegraphics[width=2.8cm]{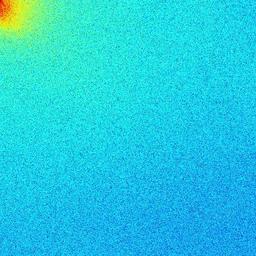}
      \includegraphics[width=2.8cm]{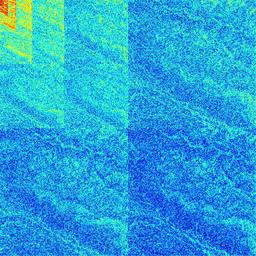}}\\
           \subfigure[CB-InSAR.]{\begin{overpic}[width=2.8cm]{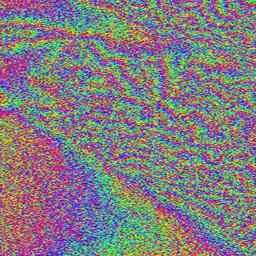}
                   \put(71.5,16){\setlength\fboxrule{1pt}\setlength{\fboxsep}{6.5pt}\color{green}\fbox{}}
          \put(1,87){\setlength\fboxrule{1pt}\setlength{\fboxsep}{6.5pt}\color{yellow}\fbox{}}
	\end{overpic}
           \begin{overpic}[width=2.8cm]{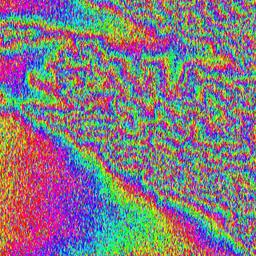}
                             \put(71.5,16){\setlength\fboxrule{1pt}\setlength{\fboxsep}{6.5pt}\color{green}\fbox{}}
          \put(1,87){\setlength\fboxrule{1pt}\setlength{\fboxsep}{6.5pt}\color{yellow}\fbox{}}
	\end{overpic}}\hspace{0.28em}
            \subfigure[NCB-InSAR with DCT]{\begin{overpic}[width=2.8cm]{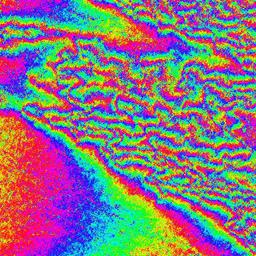}
                              \put(71.5,16){\setlength\fboxrule{1pt}\setlength{\fboxsep}{6.5pt}\color{green}\fbox{}}
          \put(1,87){\setlength\fboxrule{1pt}\setlength{\fboxsep}{6.5pt}\color{yellow}\fbox{}}
	\end{overpic}
            \begin{overpic}[width=2.8cm]{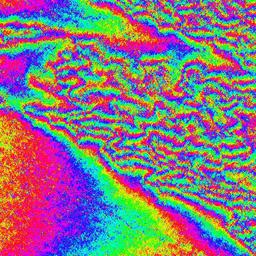}
                              \put(71.5,16){\setlength\fboxrule{1pt}\setlength{\fboxsep}{6.5pt}\color{green}\fbox{}}
          \put(1,87){\setlength\fboxrule{1pt}\setlength{\fboxsep}{6.5pt}\color{yellow}\fbox{}}
	\end{overpic}}\hspace{0.28em}
                \subfigure[NCB-InSAR with DB-4]{\begin{overpic}[width=2.8cm]{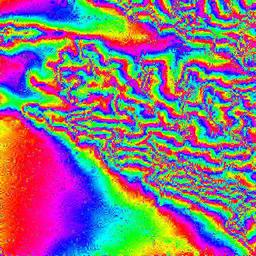}
                  \put(71.5,16){\setlength\fboxrule{1pt}\setlength{\fboxsep}{6.5pt}\color{green}\fbox{}}
          \put(1,87){\setlength\fboxrule{1pt}\setlength{\fboxsep}{6.5pt}\color{yellow}\fbox{}}
	\end{overpic}
           \begin{overpic}[width=2.8cm]{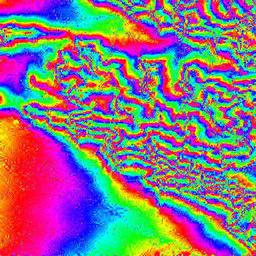}
                    \put(71.5,16){\setlength\fboxrule{1pt}\setlength{\fboxsep}{6.5pt}\color{green}\fbox{}}
          \put(1,87){\setlength\fboxrule{1pt}\setlength{\fboxsep}{6.5pt}\color{yellow}\fbox{}}
	\end{overpic}
           }\\
                             \subfigure[Yellow-framed areas in (b).]{\includegraphics[width=2.8cm]{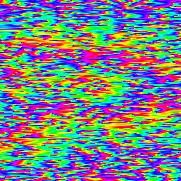}
           \includegraphics[width=2.8cm]{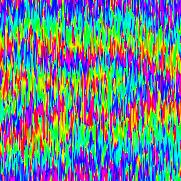}}\hspace{0.25em}
            \subfigure[Yellow-framed areas in (c).]{\includegraphics[width=2.8cm]{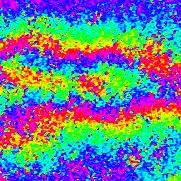}
           \includegraphics[width=2.8cm]{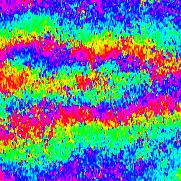}}\hspace{0.25em}
                \subfigure[Yellow-framed areas in (d).]{\includegraphics[width=2.8cm]{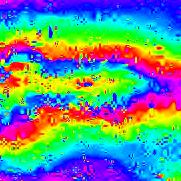}
           \includegraphics[width=2.8cm]{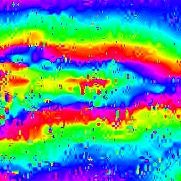}
           }\\
                    \subfigure[Green-framed areas in (b).]{\includegraphics[width=2.8cm]{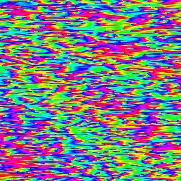}
           \includegraphics[width=2.8cm]{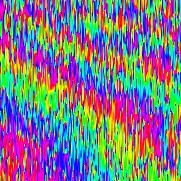}}\hspace{0.25em}
            \subfigure[Green-framed areas in (c).]{\includegraphics[width=2.8cm]{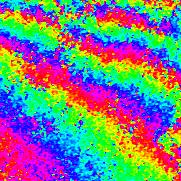}
           \includegraphics[width=2.8cm]{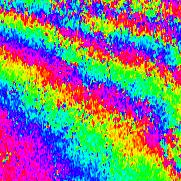}}\hspace{0.25em}
               \subfigure[Green-framed areas (d).]{ \includegraphics[width=2.8cm]{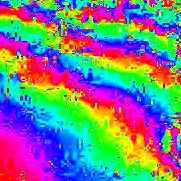}
           \includegraphics[width=2.8cm]{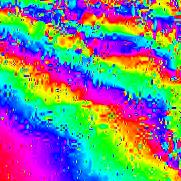}
           }\\
  \caption{ Experiments using RADARSAT-2 Phoenix dataset. (a) shows the HR reference fringes and its associated DCT and DB-4 coefficient; (b), (e) and (h) show the results of CB-InSAR; (c), (f) and (i) show the results of NCB-InSAR with DCT; (d), (g) and (j) show the results of NCB-InSAR with DB-4 wavelet. The resolution ratios in each subfigure are $1/16\times 1$ and $1\times1/16$, respectively. The range is horizontal, and the azimuth is vertical. (RADARSAT-2 Data and Products \copyright\  MDA Geospatial Services Inc. (2008) - All Right Reserved. RADARSAT is an official mark of the Canadian Space Agency.) }\label{phoenix_ifg}
\end{figure*}

\begin{figure*}
  \centering
    \includegraphics[scale=0.3]{colormap//wav_err_colormap.jpg}\\
  \subfigure[DCT coefficient error.]{\includegraphics[width=2.13cm]{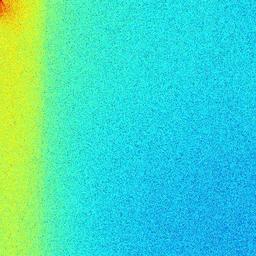}
  \includegraphics[width=2.13cm]{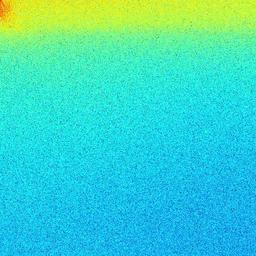}
  \includegraphics[width=2.13cm]{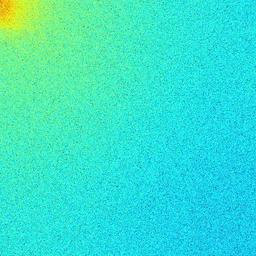}
  \includegraphics[width=2.13cm]{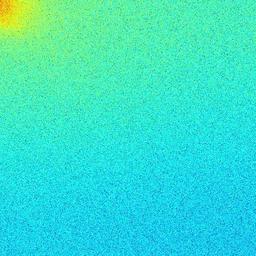}}\hspace{0.25em}
    \subfigure[DB-4 coefficient error.]{\includegraphics[width=2.13cm]{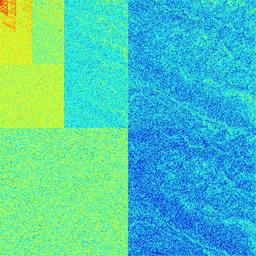}
  \includegraphics[width=2.13cm]{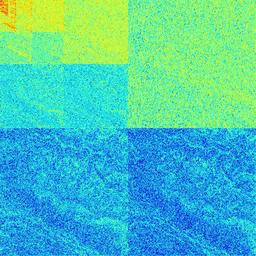}
  \includegraphics[width=2.13cm]{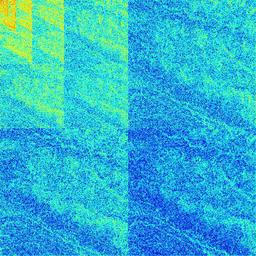}
  \includegraphics[width=2.13cm]{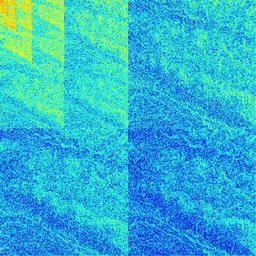}}\\
  \vspace{0.5em}
  \includegraphics[width=5cm]{colormap//legend.pdf}\\
    \subfigure[ DCT coefficient error distribution.]{\includegraphics[width=4.4cm]{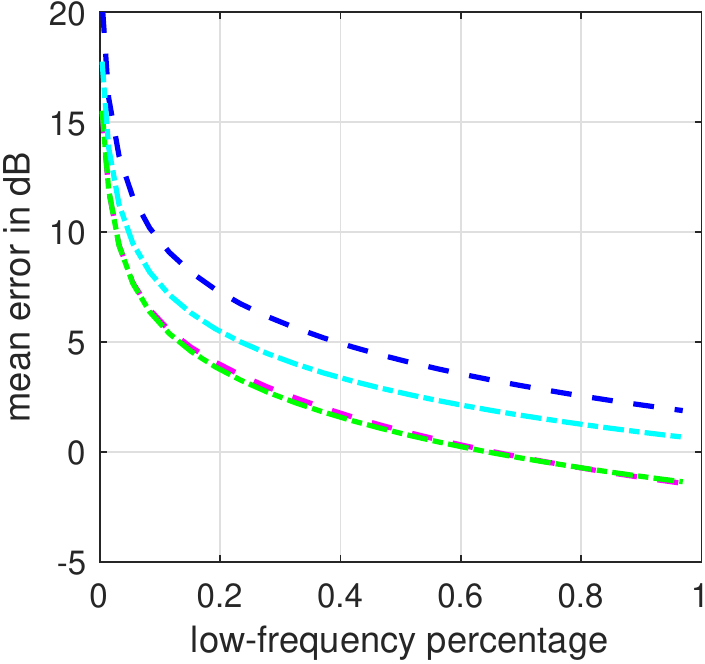}
\includegraphics[width=4.4cm]{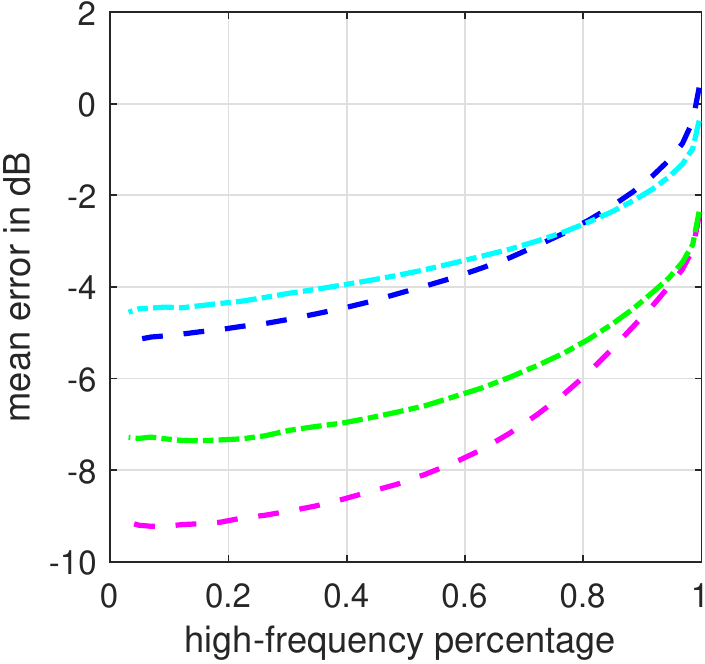}}
    \subfigure[DB-4 coefficient error distribution.]{\includegraphics[width=4.4cm]{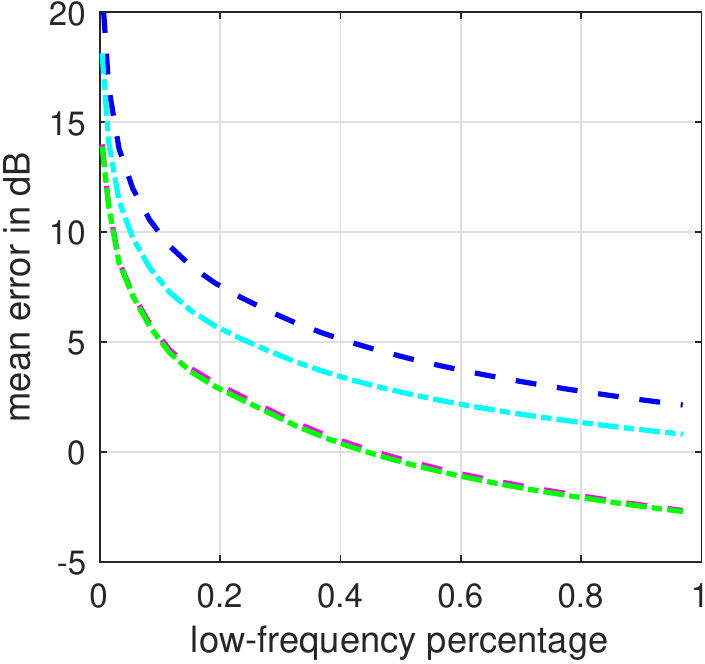}
\includegraphics[width=4.4cm]{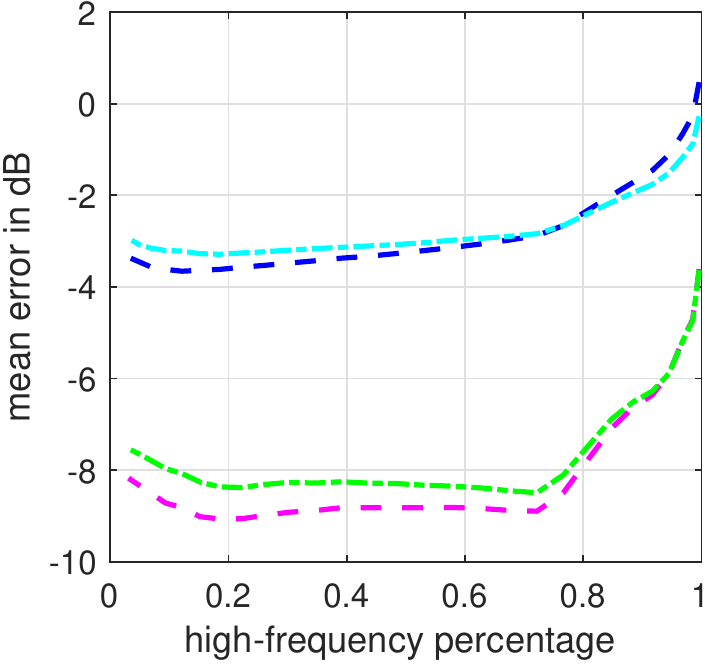}}\\
  \caption{DCT and DB-4 coefficient errors and their distributions for tests on RADARSAT-2 Phoenix Dataset. {(a)-(b) show the DCT and DB-4 coefficient errors for Figs. \ref{phoenix_ifg} (b)-(d); (c)-(d) show the corresponding error distributions.} }\label{real_freq_err}
\end{figure*}

\section{Numerical Experiments}
We test the performance of the proposed NCB-InSAR method through several numerical experiments on simulated and half-real datasets.
Several pre-processing steps are applied to the datasets to generate simulated and half-real NCB InSAR image pairs with various resolution ratios.  Then we choose the DB-4 wavelet and DCT as the sparse transforms and use the NIL1M
algorithm to recover HR interferograms with $N_{iter}=200$. {As a comparison, we also provide the results of conventional CB-InSAR.} The CB-InSAR processing steps used in this section is: filtering the master image to keep only
the CB spectrum, resampling the slave image to the master pixel spacing, interferogram formation by conjugate multiplication, and removing flat-Earth phase. Detailed experimental settings, results and analyses are given below.

 \subsection{Experiments on Simulated Dataset}
We begin by ideal simulations. The image amplitude, master image phase, and topographic phases are shown in Fig. \ref{sim_ref}. Here we consider two examples of topographic phases as shown in Fig. \ref{sim_ref}(b) and (c), respectively.
{The topographic phases are simulated for considering homogeneous fringes, and several small phase patches ($16\times 16$ pixels) are used to consider the presence of outliers.}
The amplitude is generated from i.i.d.  Rayleigh distribution and is assumed to be the same for the master and the slave images. The HR slave image is filtered by a lowpass filter  $\mathcal{S}_{\alpha,\beta}$ with several settings of $\alpha\times\beta$ to generate LR slave images.

With the above simulations, we consider a noise-free setting and a noisy setting. {The noise term $\hat{\bf e}$ is simulated with i.i.d. phase noise $\phi_{noise}$ uniformly distributed in $[-\frac{\pi}{4},\frac{\pi}{4}]$, which leads to a SNR about {$3.5$ dB and  a coherence level about $0.55$}}.  Then we recover topographic phases by CB-InSAR and NCB-InSAR. The regularization parameter of NCB-InSAR are set according to (\ref{lambda}) with $\gamma=1$, and the sparse basis is DCT.

Fig. \ref{sim_noisefree} {shows} the topographic fringes recovered by CB-InSAR and NCB-InSAR in the noise-free setting with resolution ratios $1/16\times 1$ and $1\times1/16$, respectively.
As shown by the left two images in Figs. \ref{sim_noisefree}(a) and (b), CB-InSAR leads to less visible fringes {due to resolution reduction}, degrading the interferogram quality. {The fringes obtained by NCB-InSAR are more visible than CB results}, as shown by the right two images in Figs. \ref{sim_noisefree}(a) and (b). To provide a clear view for pixel phase responses, we zoom in two small areas of the fringes which are marked by yellow and green frames in Figs. \ref{sim_noisefree}(a)-(b) and displayed separately in Figs. \ref{sim_noisefree}(c)-(f).
{The zoom-in areas of CB results clearly show a phase noise effect, which is introduced by reducing SLC images' resolution. In the worst case, the fringes are completely invisible by CB interferometry, as seen from the first zoom-in areas in Fig. \ref{sim_noisefree}(e).  On the contrary, our method still preserves the fringes with less noise effect.}
{However, the small phase patches cannot be preserved by NCB-InSAR. The reason is that these small patches act like outliers in the homogeneous fringes and are not sparse enough in the DCT basis. Based on these findings, we consider that NCB-InSAR improves interferogram quality in the sense of phase noise reduction, but cannot preserve spatially sparse outliers.}
Fig. \ref{sim_noisy} shows the results in the noisy setting, which are consistent with the results in Fig. \ref{sim_noisefree}.

Figs. \ref{sim_freq_err}(a)-(d) show the DCT coefficients of exponential phase errors for Figs. \ref{sim_noisefree} (a)-(b) and Figs. \ref{sim_noisy} (a)-(b), defined by the reference phase $\phi_{topo}[n,l]$ and the recovered phase $\phi^{\#}_{topo}[n,l]$  as
\begin{equation}\label{}
  {E}_{coef}[p,q]=\mathcal{W}\left\{e^{j\phi_{topo}[n,l]}-e^{j\phi_{topo}^{\#}[n,l]}\right\},
\end{equation}
where $\mathcal{W}$ denotes the sparse transform.  In our experiments, $p,q\in\{1,\cdots, N\}$ since the images are square and $[p,q]$ and $[n,l]$ has the same size.
Due to resolution reduction of SLC images, $e^{j\phi_{topo}^{\#}[n,l]}$ recovered by CB-InSAR contains phase noise, which whiten the spectrum of $e^{j\phi_{topo}^{\#}[n,l]}$.  The joint effect of phase noise and oversampling leads to the CB-InSAR  error maps shown in the first and second images in Figs. \ref{sim_freq_err}(a)-(d). By NCB-InSAR, the mean errors of DCT coefficients are relatively smaller than the errors by CB-InSAR,  especially in the low-frequency domain. This shows that NCB-InSAR better preserves the information of  original interferogram.

To quantitatively evaluate the coefficient errors, we compute the mean coefficient errors with respect to low and high-frequency percentages, respectively. We refer to ${E}_{coef}[p,q]$ for $p,q\le P$ as the  low-frequency coefficients with  low-frequency percentage $\xi_{l}={P^2}/{N^2}$. Then we define the mean coefficient error with respect to low-frequency percentage $\xi_{l}$  as
\begin{equation}\label{}
  E_{low}(\xi_{l})=10\,\mathrm{log}_{10}\left(\frac{1}{P^2}\sum_{p,q\le P}\big|{E}_{coef}[p,q]\big|^2\right).
\end{equation}
Similarly,  we refer to ${E}_{coef}[p,q]$ for $p,q> P$ as the  high-frequency coefficients with high-frequency percentage $\xi_{h}=1-{P^2}/{N^2}$, and define the mean coefficient error with respect to high-frequency percentage $\xi_{h}$ as
\begin{equation}\label{}
  E_{high}(\xi_{h})=10\,\mathrm{log}_{10}\left(\frac{1}{N^2-P^2}\sum_{p,q> P}\big|{E}_{coef}[p,q]\big|^2\right).
\end{equation}
Taking different values of $P$ gives the error curves $E_{low}(\xi_{l})$ and $E_{high}(\xi_{h})$. The error curves for each simulation setting is shown in Figs. \ref{sim_freq_err}(e)-(l). As seen from the curves, NCB-InSAR errors are clearly less than CB-InSAR errors in both low and high-frequency domains. For example,  Figs. \ref{sim_freq_err}(f) and (j) show that, the mean errors for $50$\% low and high frequencies by NCB-InSAR are about {$10.4$ dB and $4.2$ dB} smaller than those by CB-InSAR, respectively.
\begin{table}[]
	\renewcommand{\arraystretch}{1.3}
	\caption{RMSEs [$\mathrm{rad}$] of recovered phases for  simulated datasets.}
	\label{tab1}
	\centering
	\begin{tabular}{l  l c r }
		\toprule
		\bfseries & Resolution ratio &$1/16\times1$ & $1\times 1/16 $ \\
		\hline
  \multicolumn{4}{l}{Noise-free setting} \\
		      &  Example 1, CB-InSAR        &1.4316    &1.0736     \\
 		      &  Example 2, CB-InSAR       &0.7583    &0.7544      \\
 & Example 1, NCB-InSAR       &0.2790    &0.2774   \\
    &  Example 2, NCB-InSAR     &0.3571    &0.3586       \\
		\hline
  \multicolumn{1}{l}{Noisy setting} \\
		        &  Example1 , CB-InSAR   &0.9517    &0.9489   \\
 &  Example 2, CB-InSAR       &1.0445    &1.0462 \\
 & Example 1, NCB-InSAR      &0.4136    &0.4126  \\
	          &  Example 2, NCB-InSAR    &0.6188    &0.6127     \\
                  \bottomrule
	\end{tabular}
\end{table}

Since our method solves a signal recovery problem, root-mean-square error (RMSE) can be used for evaluating the performance of our method.
The RMSE of the recovered topographic phase is defined as
\begin{equation}\label{}
	\mathrm{RMSE}=\sqrt{\mathrm{mean}\left\{\left\|\mathrm{angle} \left\{e^{j{\phi_{topo}[n,l]-j\phi_{topo}^{\#}[n,l]}}\right\}\right\|_2^2\right\}},
\end{equation}
where $\mathrm{angle}\{\cdot\}$ and $\mathrm{mean}\{\cdot\}$ take the angle and mean values of the bracketed variables, respectively.
As shown in Table \uppercase\expandafter{\romannumeral1}, NCB-InSAR has significantly smaller RMSEs than CB-InSAR.
For example, when the resolution ratio is $1\times 1/16$, the RMSEs of CB-InSAR results are above $0.94$ [rad] in the noisy setting, while NCB-InSAR only leads to about {$0.41$ [rad] and $0.61$ [rad]} RMSEs for the two simulated examples, respectively.
These results reveal that NCB-InSAR can effectively recover high-quality interferograms compared to CB-InSAR for processing the NCB InSAR image pairs.

We should point out that the above simulations of simple geometrical fringes are ideal, considering that in practice interferometric fringes usually assume more complex patterns. Besides, scattering phase seen by different acquisitions can be decorrelated and the noise $\phi_{noise}$ may not obey an i.i.d. uniform distribution. To better test our method, we next perform numerical experiments based on a real SAR dataset owing to its complex interferometric fringes.

\subsection{Experiments on RADARSAT-2 Phoenix Dataset}
The Phoenix dataset is an ultra-fine interferometric pair collected by RADARSAT-2 on 4 and 28, May, 2008, respectively. The scene is located in Phoenix,  Alabama, USA. From this dataset we extract a subset of $1024\times 1024$ pixels for our experiments. The pixel spacing is $1.3\times2.1$ [m] with oversampling factors $1.1$ in range and $1.3$ in azimuth. Similar to previous settings, we generate half-real NCB InSAR image pairs by lowpass filtering one of original images. The regularization parameter is chosen by (\ref{lambda}) with $\gamma=0.25$.

\begin{table}[]
	\renewcommand{\arraystretch}{1.3}
	\caption{RMSEs [$\mathrm{rad}$] of recovered phases for RADARSAT-2 datasets}
	\label{tab1}
	\centering
	\begin{tabular}{l  l c r }
		\toprule
		\bfseries & Resolution ratio & $1/16\times1$ & $1 \times 1/16$ \\
		\hline
    &   CB-InSAR                &1.5296         &1.3088       \\
 	&   NCB-InSAR, DCT     &0.9815         &0.9931       \\
    &   NCB-InSAR, DB-4    &0.8446         &0.8433       \\
		\bottomrule
	\end{tabular}
\end{table}
{The results on this dataset  is shown in Fig. \ref{phoenix_ifg}. Similar to the previous results, NCB-InSAR better recovers the fringes while CB-InSAR suffers from the noise effect due to resolution reduction of SAR images. This point is clearly seen from the zoom-in areas: in Figs. \ref{phoenix_ifg}(e) and (h), the fringes processed by CB-InSAR is hardly visible, whereas in Figs. \ref{phoenix_ifg}(f), (g), (i) and (j), the fringes processed by NCB-InSAR is still clearly visible. This means an improvement in interferogram quality, as one can easily find from the pixel responses in the zoom-in areas. For providing a simple numerical comparison, we use a filtered version (using Goldstein filtering) of the original HR fringes as a reference to compute DCT/wavelet errors and RMSEs. As the results shown in Fig. \ref{real_freq_err}(d), by NCB-InSAR, the mean  errors of $50\%$ high and low-frequency DB-4 wavelet coefficients are approximately $3$ dB smaller than those by CB-InSAR. RMSEs are also reduced by $0.3\sim 0.7$ [rad], as shown in Table  \uppercase\expandafter{\romannumeral2}.}

Regarding the sparse transforms used in NCB-InSAR, DB-4 wavelet leads to better results than DCT for  this datasets. Specifically,  the RMSEs by DB-4 wavelet is about $0.15$ [rad] smaller  than those by DCT.
This suggests that the choice of a proper sparse basis is beneficial for the retrieval of interferograms. How to choose better sparse bases will be considered in our future works.

{\subsection{Spectral Extrapolation}
We find that the proposed method can generate extrapolated interferogram spectra with respect to CB processing.
To show this point, we provide the image spectra and interferogram spectra of Figs. 9 and 11 in Figs. \ref{sim_spec} and \ref{radarsat_spec}, respectively. For our method, we have multiplied the output of Algorithm 1 by the master image amplitude to adapt our interferograms to the conventional definition of interferograms, $|z_m[n,l]|\cdot |z_s[n,l]|e^{j\phi_{topo}}=|z_m[n,l]|\cdot \overline{u}[n,l]$.}

{To facilitate analysis, we provide the theoretical description of interferogram spectra. By the convolution theorem, the interferogram spectra obtained by spatial-domain conjugate multiplication can be expressed in frequency-domain by
\begin{equation}\label{}
  I(f_r,f_a)=\iint \overline{Z}_s(\nu -f_r,\xi -f_a)Z_m(\nu,\xi)d\nu d\xi.
\end{equation}}

{For the CB image pair that has a partial spectral band, the image spectra outside the CB are discarded, and it therefore brings a loss of interferometric information, which can be clearly shown by the convolution equation. Accordingly, the LR CB interferogram spectra get weaker and narrower due to the discarding of coherent spectra, as shown by the presented Figs. \ref{sim_spec} and \ref{radarsat_spec}.}
{In contrast to CB processing, the interferogram spectra produced by the proposed method are close to the original ones. This coincides with the aim of our method which tries to recover the original HR interferograms. However, the original interferograms are not completely recovered, since the spectral extrapolation is not perfect. Specifically, the spectra recovered by our method are cleaner and weaker than the original ones, so we cannot achieve the same resolution as the original one.}

{It is noteworthy that the achievement of spectra extrapolation is signal-dependent. For geometrically simple cases, like the simulated examples, the spectra can be well extrapolated; for complicated cases like the presented RADARSAT-2 example, spectral extrapolation is not a easy task but can still be obtained to a noticeable degree with respect to CB processing.}

\begin{figure*}
  \centering
\subfigure[Example 1.]{\includegraphics[width=2.1cm]{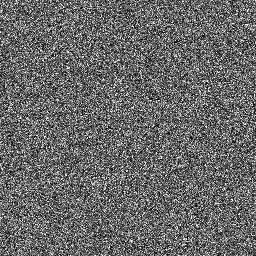}
\includegraphics[width=2.1cm]{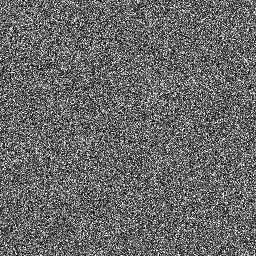}
\includegraphics[width=2.1cm]{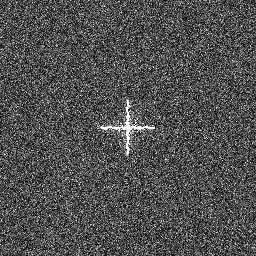}}\hspace{2.5cm}
\subfigure[Example 2.]{\includegraphics[width=2.1cm]{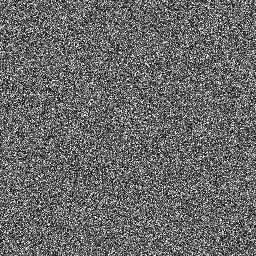}
\includegraphics[width=2.1cm]{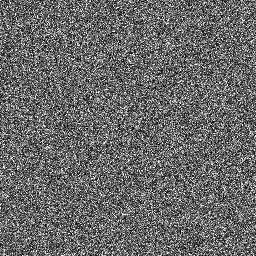}
\includegraphics[width=2.1cm]{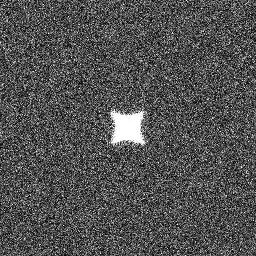}}\\
\subfigure[$1/16\times 1$, example 1.]{\includegraphics[width=2.1cm]{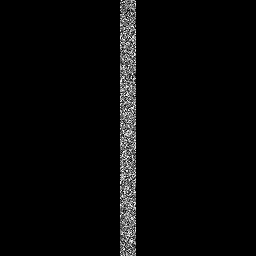}
\includegraphics[width=2.1cm]{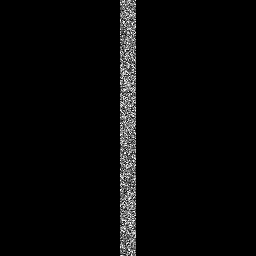}
\includegraphics[width=2.1cm]{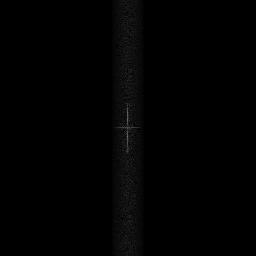}
\includegraphics[width=2.1cm]{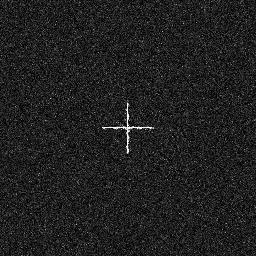}}\hspace{0.2cm}
\subfigure[$1/16\times 1$, example 2.]{\includegraphics[width=2.1cm]{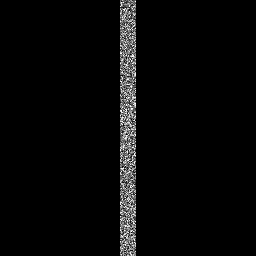}
\includegraphics[width=2.1cm]{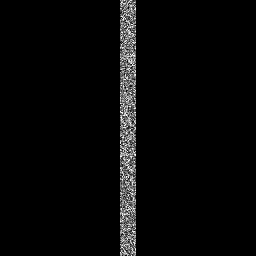}
\includegraphics[width=2.1cm]{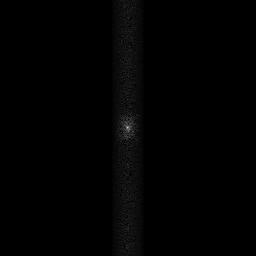}
\includegraphics[width=2.1cm]{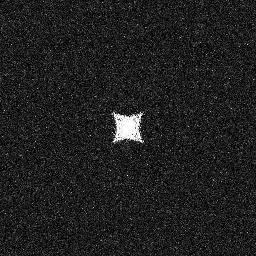}}\\
\subfigure[$1\times 1/16$, example 1.]{\includegraphics[width=2.1cm]{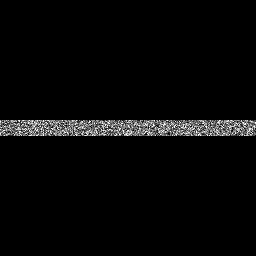}
\includegraphics[width=2.1cm]{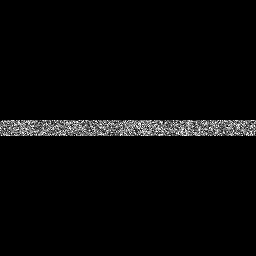}
\includegraphics[width=2.1cm]{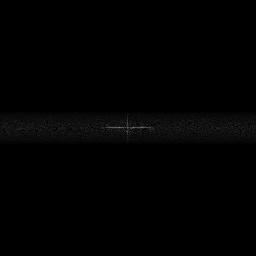}
\includegraphics[width=2.1cm]{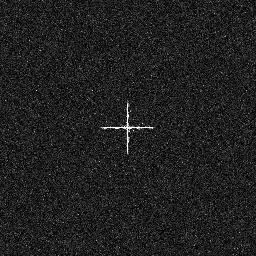}}\hspace{0.2cm}
\subfigure[$1\times 1/16$, example 2.]{\includegraphics[width=2.1cm]{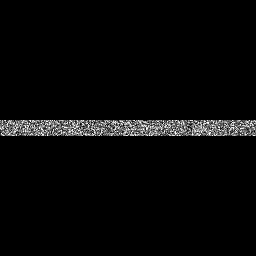}
\includegraphics[width=2.1cm]{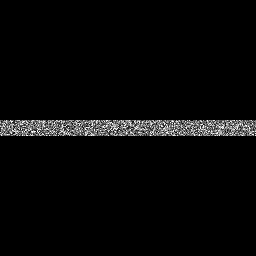}
\includegraphics[width=2.1cm]{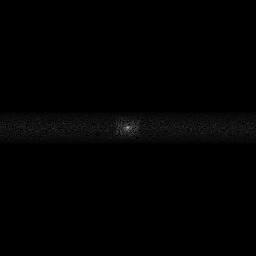}
\includegraphics[width=2.1cm]{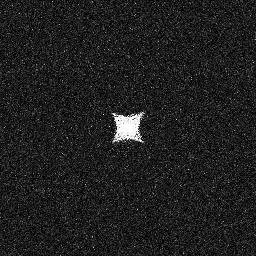}}\\
  \caption{Spectra of simulated images and interferograms in the noisy setting. For each subfigure of (a)-(b): master spectrum, slave spectrum, HR interferogram spectrum. For each subfigure of (c)-(f): LR master spectrum, LR slave spectrum, CB interferogram spectrum, NCB-InSAR interferogram spectrum. Black-to-white colormap range: $[0, 2.20\mathrm{E}3]$ for image spectra, $[0, 3.20\mathrm{E}3]$ for interferogram spectra. }\label{sim_spec}
\end{figure*}

\begin{figure*}
  \centering
\subfigure[]{\includegraphics[width=2.2cm]{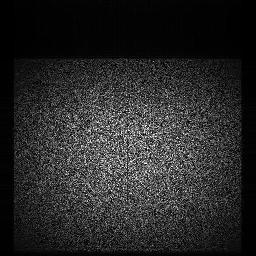}
\includegraphics[width=2.2cm]{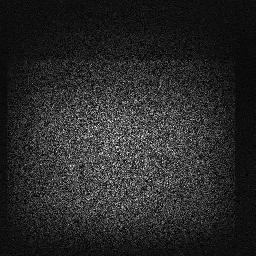}
\includegraphics[width=2.2cm]{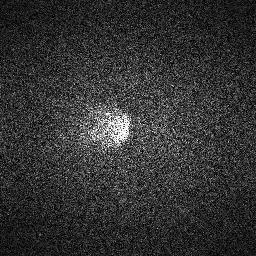}}
\subfigure[$1/16\times 1$.]{\includegraphics[width=2.2cm]{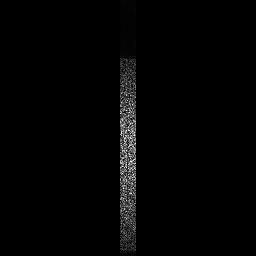}
\includegraphics[width=2.2cm]{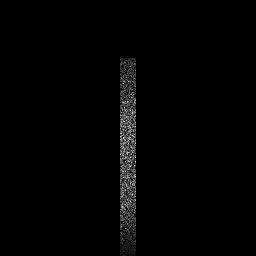}
\includegraphics[width=2.2cm]{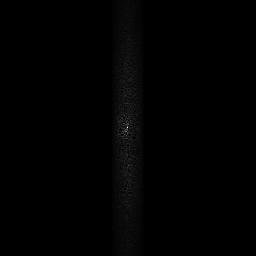}
\includegraphics[width=2.2cm]{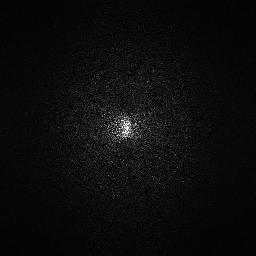}
\includegraphics[width=2.2cm]{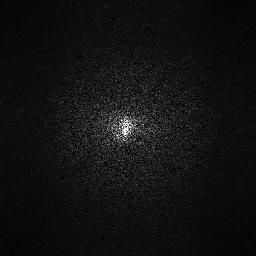}}\hspace{0.2cm}
\subfigure[$1\times 1/16$.]{\includegraphics[width=2.2cm]{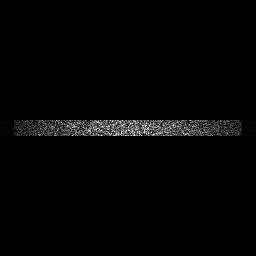}
\includegraphics[width=2.2cm]{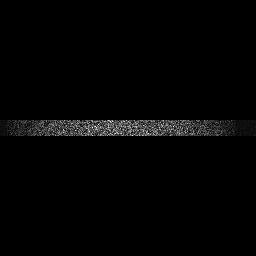}
\includegraphics[width=2.2cm]{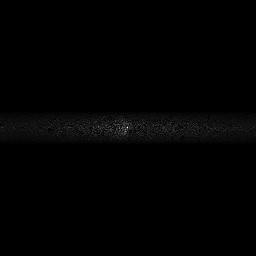}
\includegraphics[width=2.2cm]{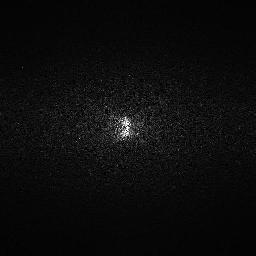}
\includegraphics[width=2.2cm]{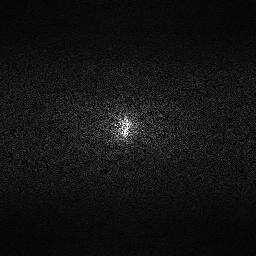}}\\
  \caption{ Spectra of RADARSAT-2 images and interferograms. For each subfigure of (a): master spectrum, slave spectrum, HR interferogram spectrum. For each subfigure of (b)-(c): LR master spectrum, LR slave spectrum, CB interferogram spectrum, NCB-InSAR interferogram spectra using DCT and DB-4 wavelet respectively.  Black-to-white colormap range: $[0, 4.67\mathrm{E}5]$ for image spectra, $[0, 7.45\mathrm{E}7]$ for interferogram spectra. }\label{radarsat_spec}
\end{figure*}

\subsection{Discussions}
The resolution reduction in the SLC image domain introduces an unwanted phase noise effect.
{The reason for this effect is that, when the image resolution gets coarser, phase variation of scatterers within a resolution cell increases. Consequently, it produces spatial decorrelation noise due to the sinc blurring kernel in the image domain. Our method is designed to recover the original fringes by solving a sparse recovery problem, which can suppress the unwanted noise effect introduced by the resolution reduction. In this sense, our method can be viewed as a special phase denoising method.} On the other hand, our method also can be viewed as a spectral extrapolation method, which is based on the exploitation of the NCB information.

We should point out that the details of our recovered fringes are not as fine as those of the original HR ones, which is seen from Figs. \ref{sim_noisefree}, \ref{sim_noisy} and \ref{phoenix_ifg}. In fact, perfect recovery cannot be achieved, because there always exists measurement noise and interferograms are not ideally sparse. Moreover, we also point out that our method requires more computational cost. Specifically, {in the case of $1024\times 1024$ pixel size the runtime (on Intel i7-7700 CPU, 3.6 [GHz], $200$ iterations) is about $163.36$ [s] for NCB-InSAR and $0.09$ [s] for CB-InSAR.}

\section{Conclusion}
This paper proposed a non-common band InSAR method in the framework of CS. In contrast to conventional CB-InSAR, the proposed NCB-InSAR exploits the full spectra of two interferometric images to perform spectral extrapolation and improve the interferogram quality.
The main characteristics and advantages of NCB-InSAR are summarized as follows:
\begin{enumerate}
  \item Compared to CB-InSAR, NCB-InSAR improves the interferogram quality by means of exploiting the non-common band.
  \item NCB-InSAR adopts CS as a signal processing approach to recover interferograms, which is different from the conventional interferometry via pixel-by-pixel conjugate multiplication.
  \item Due to the speckle effect, the CS model of NCB-InSAR naturally satisfies the RIP requirement in CS. As no additional random sub-sampling strategy is required, NCB-InSAR can be directly applied to existing spaceborne and airborne SAR systems.
  \item In NCB-InSAR, the resolution of the slave image can be much coarser than that of the master one. This suggests a  reduction in the cost and requirement on data acquisition.
\end{enumerate}
For computational efficiency, we also designed a specialized algorithm for NCB-InSAR. The algorithm mainly involves fast matrix operations, like FFT and wavelet transfrom, and hence is computationally efficient.

{We would like to point out that the proposed method is not a universal one. Limited by the CS theory, it is designed for specific cases, i.e., extended targets having homogeneous fringes. It is not suitable for recovering interferometric phases of point-like scatterers or outlier patches, and therefore is not applicable for urban areas.}

{Currently, our experiments are limited to simulated and half-real datasets. In future works, we would like to seek real InSAR datasets acquired with different resolutions, and investigate how well our method works on real data in the presence of error sources  like non-ideal antenna weighting, coregistration error and decorrelation. Moreover, since NCB-InSAR requires a HR image, we would like to further study the proposed method in HR case like spotlight InSAR \cite{Eineder2009,Ge2018}, and evaluate the performance of NCB-InSAR on HR datasets.}

\section*{Acknowledgement}
{The authors would like to thank NASA/JPL-Caltech, European Space Agency, MDA
Geospatial Services Inc. and Canadian Space Agency for providing open SAR data. Thank ESA's SAR processing software SNAP and NEST, which are important tools during our work.
{Thank the anonymous reviewers for valuable suggestions}, which helps improving this paper.}

\begin{appendices}
	\section{Proof of Lemma 1 }

	The proof exploits the CoM phenomenon for Lipschitz functions of random variables\cite{Tao2012Topics}. {In the following proof, we rely on the assumption that each entry of $\bm{\theta}_m$ is independently uniformly distributed on the unit circle of ${\mathbb C}$.}

	Define a function $f(\bm{\theta}_m)=\norm{\mathbf{H}_m\mathbf{u}}_2$. The first step of our proof is to derive the Lipschitz constant of the function.
	Referring to (\ref{rd0}) and (\ref{rd2}), it  can be derived that
	\begin{equation}\label{}
		\begin{split}
			f(\bm{\theta}_m)&=\norml{{{\mathbf{H}}_m}\mathbf{u} }_2\\
			&=\frac{1}{\sqrt{\alpha\beta}}\norm{(\mathbf{F}_{J}^*\otimes\mathbf{F}_{I}^*)\mathbf{\Omega}_{2D} \mathbf{F}_{2D}\mathrm{diag}({\bm\theta}_m)\bf u}_2\\
			&=\frac{1}{\sqrt{\alpha\beta}}\norm{\mathbf{\Omega}_{2D} \mathbf{F}_{2D}\mathrm{diag}( \mathbf{u})\bm{\theta}_m}_2\\
			&=\frac{1}{\sqrt{\alpha\beta}}\norm{\mathbf{\Omega}_{2D} \mathbf{F}_{2D}\mathrm{diag}(\mathbf{u})\mathbf{F}_{2D}^*\mathbf{F}_{2D}\bm{\theta}_m}_2\\
			&=\frac{1}{\sqrt{\alpha\beta}}\norm{\mathbf{\Omega}_{2D}}_2\norm{\mathbf{F}_{2D}\mathrm{diag}(\mathbf{u})\mathbf{F}_{2D}^*}_2\norm{\bm{\theta}_m}_2\\
			&\le\frac{1}{\sqrt{\alpha\beta}}\norm{\mathbf{u}}_{\infty}\norm{\bm{\theta}_m}_2.\\
		\end{split}
	\end{equation}
	Without loss of generality, we first consider the case of $\norm{\bf u}_2=1$. Then the function $f(\bm{\theta}_m)$ has a  Lipschitz constant $\frac{\sigma_{\bf u}}{\sqrt{\alpha\beta}}$, where
	\begin{equation}\label{}
		\sigma_{\bf u}=\sup\limits_{\|{\bf u}\|_2=1}\norm{\mathbf{u}}_{\infty}
	\end{equation}

	Secondly, we use the {Talagrand concentration inequality} (Proposition 1.1.13 of \cite{Tao2012Topics}) to derive the CoM inequality for the Lipschitz function $f(\bm{\theta}_m)$.

	\emph{Lemma 2 (Talagrand concentration inequality):} Let $B > 0$, and let
	$\bm{\zeta}=[\zeta_1,\cdots,\zeta_K]$ be a variable with independent entries $\left|\zeta_k\right|\le B$ for all $1\le p\le K$. Let $f: \mathbb{C}^K \rightarrow \mathbb{R}$ be a $\sigma$-Lipschitz convex function. Then for any $\delta$, one has
	\begin{equation}\label{a10b1}
		\mathbb{P}\left\{\left|f(\bm\zeta)-\mathbb{M}f(\bm\zeta)\right|\ge\delta\right\}\le C_3 \mathrm{exp}\left(-\frac{C_4\delta^2}{\sigma^2B^2}\right),
	\end{equation}
	where $\mathbb{P}\{\cdot\}$ denotes the probability of an event,  $\mathbb{M}f(\bm\zeta)$ is the probabilistic median of $f(\bm\zeta)$, and  {$C_3=4$,} $C_4=1/16$ are universal positive constants \cite{Massart2000About}.

	For the Lipschitz function $f(\bm{\theta}_m)$, the variable $\bm{\theta}_m$ has i.i.d. bounded entries $\left|\theta_{m,n}\right|\le 1$ for $n=1,\cdots,K$, where $\theta_{m,n}$ denotes the $n$-th entry of ${\bm\theta}_{m}$. Therefore, by Lemma 2 we have
	\begin{equation}\label{}
		\mathbb{P}\left\{\left |f(\mathbf{\bm{\theta}}_m)-\mathbb{M}f({\bf \bm{\theta}}_m)\right|\ge\delta\right\}\le4\exp \left( -\frac{{{\alpha\beta\delta }^{2}}}{16{{\sigma_{\bf u}^2}}} \right).
	\end{equation}
	This gives
	\begin{equation}\label{a50}
		\mathbb{P}\left\{\left |\norml{{\mathbf{H}}_m{\mathbf{u}}}_2-\mathbb{M}f({\bf \bm{\theta}}_m)\right|\ge\delta\right\}\le4\exp \left( -\frac{{{\alpha\beta\delta }^{2}}}{16{{\sigma_{\bf u}^2}}} \right).
	\end{equation}

	Next, we are going to derive the median $\mathbb{M}f({\bf \bm{\theta}}_m)$. Denoting
	\begin{equation}\label{}
		\mathbf{T}_{\bf u}=\frac{1}{\sqrt{\alpha\beta}}\mathbf{\Omega}_{2D} \mathbf{F}_{2D}\mathrm{diag}( \mathbf{u}),
	\end{equation}
	and $\mathbf{Q}=\mathbf{T}_{\bf u}^*\mathbf{T}_{\bf u}$,
	the square of random function $f(\bm{\theta}_m)$ can be written as
	\begin{equation}\label{sym}
		\begin{split}
			f^2(\bm{\theta}_m)&=\norm{\mathbf{T}\bm{\theta}_m}_2^2
			=\bm{\theta}_m^*\mathbf{Q}\bm{\theta}_m\\
			&=\sum\limits_{i\neq j}{\theta}_{m,i}^* Q_{i,j}{\theta}_{m,j}+\sum\limits_{i}{\theta}_{m,i}^* Q_{i,i}{\theta}_{m,i}\\
			&=\sum\limits_{i< j}2\mathrm{Re}\left\{{\theta}_{m,i}^* Q_{i,j}{\theta}_{m,j}\right\}+1.
		\end{split}
	\end{equation}
	The last equality holds because $Q_{i,j}=Q_{j,i}^*$ for $i\neq j$, and $Q_{i,i}=|u_i|^2$. From (\ref{sym}), we can see that the function $f^2(\bm{\theta}_m)$ is symmetric about $f^2(\bm{\theta}_m)=1$ {because the first term in the last line take negative and positive values with equal probability}. So it is easy to find that the events $\{f^2(\bm{\theta}_m)>1\}$ and $\{f^2(\bm{\theta}_m)<1\}$ have the same probability.
	It then follows $\mathbb{M}f^2(\bm{\theta}_m)=1$. Consequently, this gives $\mathbb{M}f(\bm{\theta}_m)=1$, since the map between  $f^2(\bm{\theta}_m)$ and $f(\bm{\theta}_m)$ is one-to-one.
	
Now we have
	\begin{equation}\label{a5}
		\begin{split}
			\mathbb{P}\left\{\left |\norml{{\mathbf{H}}_m\mathbf{u}}_2^2-1\right|\ge\delta\right\}
		\le4\exp \left( -\frac{{{\alpha\beta\delta^2 }}}{16{{\sigma_{\bf u} }^{2}}} \right).
		\end{split}
	\end{equation}

	For an arbitrary vector $\bf u$, its normalized version $\frac{\mathbf{u}}{\|{\mathbf{u}}\|_2}$ has a unit norm, so
	\begin{equation}
		\begin{split}
			&\mathbb{P}\left\{\left |\norml{{\mathbf{H}}_m\mathbf{u}}_2-\norm{\bf u}_2\right|\ge\delta\norm{\bf u}_2\right\}\\
			&=\mathbb{P}\left\{\left |\left\|{{\mathbf{H}}_m\frac{\mathbf{u}}{\norm{\bf u}_2}}\right\|_2-1\right|\ge\delta\right\}\\
			&\le4\exp \left( -\frac{{{\alpha\beta\delta^2 }}}{16{{\sigma_{\bf u} }^{2}}} \right).
		\end{split}
	\end{equation}
	Since $\bf u=Wx$, for $k$-sparse vector $\bf x$, we have
	\begin{equation}\label{}
		\begin{split}
			\sigma_{\bf u} &= \sup\limits_{\left\|{\bf Wx}\right\|_2=1,|\mathrm{supp}(\mathbf{x})|\le k}{{\|{\bf Wx} \|}_{\infty}} \\
			&= \sup\limits_{\left\|{\bf x}\right\|_2=1,|\mathrm{supp}(\mathbf{x})|\le k} {{\norm{\bf Wx} }_{\infty}}\\
			&=\norm{\mathbf{W}}_{k,\infty}.
		\end{split}
	\end{equation}

	Finally, considering that $\norml{{\mathbf{H}}_m\mathbf{u}}_2=\norm{\mathbf{A}_m\mathbf{x}}_2$, and $\norm{\bf u}_2=\norm{\bf x}_2$ {due to the orthonormality of $\bf W$},  we have
	\begin{equation}\label{}
		\begin{split}
			\mathbb{P}\left\{\big |\norm{\mathbf{A}_m\mathbf{x}}_2-\norm{\bf x}_2\big|\ge{\delta\norm{\bf x}_2}\right\}
			\le4\exp \left( -\frac{{{\alpha\beta\delta^2 }}}{{16\norm{\mathbf{W}}_{k,\infty}^2}} \right)
		\end{split}
	\end{equation}
	with $\delta\in(0,1)$ for arbitrary $k$-sparse vector $\bf x$.
	\section{Proof of Theorem 1}
Our proof is rooted in \cite{Baraniuk2008}, which use the concentration of measure inequality and covering number estimates to provide a simple proof. Some essential modifications and extensions are considered to adapt the results to our problem.

Without loss of generality, we prove Theorem 1 in the case of $\norm{\bf x}_2=1$. First, define a $k$-dimensional subspace $X_S=\{{\bf x}| \mathrm{supp}({\bf x})=S\}$, where $S\subset \{1,\cdots,K\}$ and $|S|=k$.  Choose a finite point set $Q_S\subset X_S$ with $\norm{\bf q}_2=1$ for all ${\bf q}\in Q_S$, such that
\begin{equation}\label{a4e2}
  \min\limits_{{\bf q}\in Q_S}\norm{\mathbf{x}-\mathbf{q}}_2\le\frac{\rho}{4}.
\end{equation}
As shown in Chapter 13 of \cite{lorentz1996constructive}, there exists such a set with number of elements less than $\left(\frac{12}{\rho}\right)^k$.
For all elements of $Q_S$, we apply the concentration inequality in Lemma 1 with a setting $\delta=\frac{\rho}{2}$ to obtain
\begin{equation}\label{a4e3}
  \left(1-\frac{\rho}{2}\right)\norm{\bf q}_2\le\norm{\bf Aq}_2\le \left(1+\frac{\rho}{2}\right)\norm{\bf q}_2, \quad \mathrm{for\; all\;} {\bf q}\in Q_S.
\end{equation}
By Lemma 1 this inequality hold with probability exceeding
\begin{equation}\label{a4e4}
  1-4\left(\frac{12}{\rho}\right)^k\exp\left(-\frac{\alpha\beta\rho^2}{64 \norm{\bf W}^2_{k,\infty}}\right).
\end{equation}

Next, let $\xi$ be the smallest number such that
\begin{equation}\label{a4e5}
  \norm{\bf Ax}_2\le(1+\xi)\norm{\bf x}_2, \quad \mathrm{for\; all\;} {\bf x}\in X_S.
\end{equation}
Now we are going to show that $\xi\le\rho$.
Since $\norm{\mathbf{x}-\mathbf{q}}_2\le\frac{\rho}{4}$ for a given ${\bf q}\in Q_S$, we have
\begin{equation}\label{a4e6}
  \norm{\bf Ax}_2\le\norm{\bf Aq}_2+\norm{\bf A(x-q)}_2\le 1+\frac{\rho}{2}+(1+\xi)\frac{\rho}{4}.
\end{equation}
Since $\xi$ is defined as the smallest number satisfying (\ref{a4e5}), we have
$\xi\le \frac{\rho}{2}+(1+\xi)\frac{\rho}{4}$
which gives
\begin{equation}\label{a4e7}
\xi\le\frac{3\rho}{4}(1-\frac{\rho}{4})\le\rho.
\end{equation}
 This leads to the upper bound $\norm{\bf Ax}_2\le(1+\rho)\norm{\bf x}_2$ for all ${\bf x}\in X_S$. Similarly, we can also derive the lower bound $\norm{\bf Ax}_2\ge(1-\rho)\norm{\bf x}_2$ for all ${\bf x}\in X_S$. Thus, we have
\begin{equation}\label{a4e8}
  (1-\rho)\norm{\bf x}_2\le\norm{\bf Ax}_2\le(1+\rho)\norm{\bf x}_2, \quad \mathrm{for\; all\;} {\bf x}\in X_S
\end{equation}
with probability exceeding (\ref{a4e4}). Consider that the number of $k$-dimensional subspaces satisfies ${\binom K k}\le\left(\frac{eK}{k}\right)^k$. Here ${\binom K k}$ is a combination number.
Then taking the union of (\ref{a4e8}) over all subspaces gives
\begin{equation}\label{a4e9}
   (1-\rho)\norm{\bf x}_2\le\norm{\bf Ax}_2\le(1+\rho)\norm{\bf x}_2
\end{equation}
for all $\bf x$, $|\mathrm{supp}({\bf x})|=k$  with probability exceeding
\begin{equation}\label{a4e10}
1-4\left(\frac{eK}{k}\right)^k\left(\frac{12}{\rho}\right)^k\exp\left(-\frac{\alpha\beta \rho^2}{64 \norm{\bf W}^2_{k,\infty}}\right).
\end{equation}

Finally, choosing $\delta=3\rho$, so that $(1-\delta)\le (1-\rho)^2$ and $(1+\delta)\ge (1+\rho)^2$ for $\rho\in(0,1)$, we have
\begin{equation}\label{}
  (1-\delta)\norm{\bf x}_2^2\le\norm{\bf Ax}_2^2\le(1+\delta)\norm{\bf x}_2^2
\end{equation}
for all $\bf x$, $|\mathrm{supp}({\bf x})|=k$  with probability exceeding
\begin{equation}\label{}
  1-4\left(\frac{eK}{k}\right)^k\left(\frac{36}{\delta}\right)^k\exp\left(-\frac{\alpha\beta \delta^2}{576 \norm{\bf W}^2_{k,\infty}}\right).
\end{equation}
	So, it is sufficient to make this probability larger than  $1-\eta$ by
	\begin{equation}\label{}
			\alpha\beta\ge  {C\norm{\mathbf{W}}_{k,\infty}^2}\delta^{-2}\left(k\,\mathrm{log}\frac{eK}{k}+k\,\mathrm{log}\frac{36}{\delta}
				+\mathrm{log}\frac{4}{\eta}\right)
	\end{equation}
where $C=576$.

\section{Derivation of  Lipschitz Constant $L_f$ }
Generally, Lipschitz constant is defined as follows: given two sets $V$ and $Y$, and a function $g(\mathbf{v}): V\rightarrow Y$, if the inequality
\begin{equation}\label{}
	\norm{g(\mathbf{v}_a)-g(\mathbf{v}_b)}\le \zeta \norm{\mathbf{v}_a-\mathbf{v}_b}
\end{equation}
holds for all $\mathbf{v}_a$ and $\mathbf{v}_b$ in $V$, then the constant $\zeta$ is referred to as a Lipschitz constant for $g(\mathbf{v})$. In our problem, the function to be discussed is the gradient
\begin{equation}\label{}
	\nabla f(\mathbf{U})=2\hat{\mathcal{H}}_m^*({{\hat{\mathbf{Y}}_s}}-\hat{\mathcal{H}}_m\mathbf{U}),
\end{equation}
and the distance is measured by the Frobenius norm $\norm{\cdot}_F$.
Then for two variables $\mathbf{U}_a$ and $\mathbf{U}_b$, defining $\mathbf{u}_a=\mathrm{vec}(\mathbf{U}_a)$ and $\mathbf{u}_b=\mathrm{vec}(\mathbf{U}_b)$, we have
\begin{equation}\label{}
	\begin{split}
		\norm{\nabla f(\mathbf{U}_a)-\nabla f(\mathbf{U}_b)}_F&=2\norml{\hat{\mathcal{H}}_m^*\hat{\mathcal{H}}_m(\mathbf{U}_a-\mathbf{U}_b)}_F\\
		&=2\norml{\mathbf{F}_{2D}^*\hat{\mathbf{H}}_m^*\hat{\mathbf{H}}_m\mathbf{F}_{2D}(\mathbf{u}_a-\mathbf{u}_b)}_2\\
		&\le2\norml{\hat{\mathbf{H}}_m^*\hat{\mathbf{H}}_m}_{2}\norm{\mathbf{u}_a-\mathbf{u}_b}_2\\
		&=2\norml{\hat{\mathbf{H}}_m^*\hat{\mathbf{H}}_m}_{2}\norm{\mathbf{U}_a-\mathbf{U}_b}_F.
	\end{split}
\end{equation}
This  inequality gives a Lipschitz constant $L_f=2\norml{\hat{\mathbf{H}}_m^*\hat{\mathbf{H}}_m}_{2}$. Referring to equation (\ref{rd2}), we have
\begin{equation}\label{}
	\begin{split}
		\hat{\mathbf{H}}_m^*\hat{\mathbf{H}}_m=&\frac{1}{{\alpha\beta}}\mathbf{F}_{2D}\mathrm{diag}(\overline{\mathbf{\Theta}}_m)\mathbf{F}_{2D}^*\mathbf{\Omega}_{2D}^T\\
		&\times\mathbf{\Omega}_{2D}\mathbf{F}_{2D}\mathrm{diag}(\mathbf{\Theta}_m)\mathbf{F}_{2D}^*.
	\end{split}
\end{equation}
Defining a unitary matrix
$\mathbf{Q}=\mathbf{F}_{2D}\mathrm{diag}(\overline{\mathbf{\Theta}}_m)\mathbf{F}_{2D}^*$,
and a diagonal matrix
$\mathbf{\Sigma}=\frac{1}{\alpha\beta}\mathbf{\Omega}_{2D}^T\mathbf{\Omega}_{2D}$,
we have the eigendecomposition
\begin{equation}\label{}
	\hat{\mathbf{H}}_m^*\hat{\mathbf{H}}_m=\mathbf{Q}\mathbf{\Sigma} \mathbf{Q}^{-1}.
\end{equation}
This gives $\norml{\hat{\mathbf{H}}_m^*\hat{\mathbf{H}}_m}_{2}=1/\alpha\beta$, since all the nonzero diagonal values of $\mathbf{\Sigma}$ are $1/\alpha\beta$. Consequently, we have $L_f=2/\alpha\beta$.

\end{appendices}

\bibliographystyle{IEEEtran}
\bibliography{mybib}

\begin{IEEEbiography}[{\includegraphics[width=1in,height=1.25in,clip,keepaspectratio]{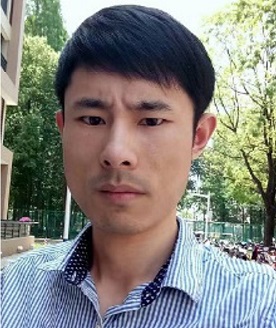}}]{Huizhang Yang}
 received the B.S. degree in communication engineering from Nanjing University of Science and Technology, Nanjing, China, in 2014. He is currently working toward the Ph.D. degree at the School of Electronic and Optical Engineering, Nanjing University of Science and Technology, Nanjing, China.

His main research interests include SAR/InSAR signal processing, sparse signal processing and compressive sensing.
\end{IEEEbiography}
 \vspace{-1.5cm}
\begin{IEEEbiography}[{\includegraphics[width=1in,height=1.25in,clip,keepaspectratio]{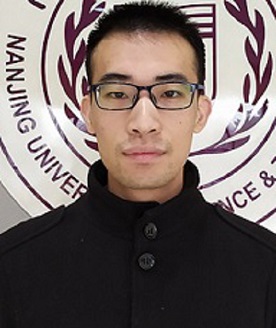}}]{Chengzhi Chen}
 received the B.S. degree in electronic information engineering from Nanjing University of Science and Technology, Nanjing, China, in 2016. He is currently working toward the Ph.D. degree at the School of Electronic and Optical Engineering, Nanjing University of Science and Technology, Nanjing, China.

His main research interests include SAR signal processing, jamming suppression, sparse signal processing and compressive sensing.
\end{IEEEbiography}
 \vspace{-1.5cm}
\begin{IEEEbiography}[{\includegraphics[width=1in,height=1.25in,clip,keepaspectratio]{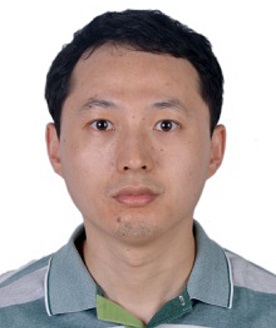}}]{Shengyao Chen}
received the B.S. degree in communication engineering and the Ph.D. degree in information and communication engineering from the Nanjing University of Science and Technology, Nanjing, China, in 2006 and 2013, respectively.

From 2013 to 2015, he was a Postdoctoral Researcher with the School of Electronic and Optical Engineering, Nanjing University of Science and Technology. He is currently an Associate Professor with the School of Electronic and Optical Engineering, Nanjing University of Science and Technology, Nanjing, China. His research interests include radar signal processing, compressive sensing, and chaotic dynamical systems.
\end{IEEEbiography}
 \vspace{-1.5cm}
\begin{IEEEbiography}[{\includegraphics[width=1in,height=1.25in,clip,keepaspectratio]{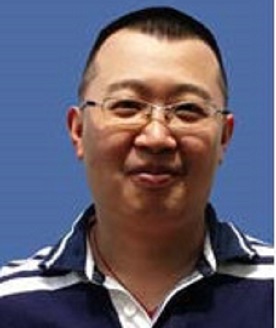}}]{Feng Xi}
received the B.S. degree in electronic engineering and the Ph.D. degree in information and communication engineering from the Nanjing University of Science and Technology, Nanjing, China, in 2003 and 2010, respectively.

Currently, he is an Associate Professor with the Department of Electronic Engineering, Nanjing University of Science and Technology, Nanjing, China. His research interests include radar signal processing, compressive sensing, and wireless sensor networks.
\end{IEEEbiography}
 \vspace{-1.5cm}
\begin{IEEEbiography}[{\includegraphics[width=1in,height=1.25in,clip,keepaspectratio]{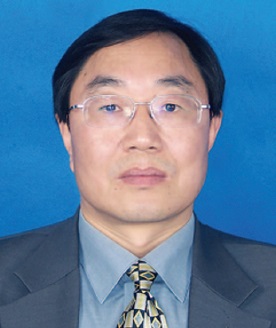}}]{Zhong Liu}
 (M¡¯03) received the B.S. degree from Anhui University, Hefei, China, in 1983 and the M.S. and Ph.D. degrees from the University of Electronic Science and Technology of China, Chengdu, China, in 1985 and 1988, all in electrical engineering.

Since 1989, he has been with the Department of Electronic Engineering, Nanjing University of Science and Technology, Nanjing, China, where he is currently a professor. He was a Postdoctoral Fellow at Kyoto University, Kyoto, Japan, from 1991 to 1993. He was a Researcher at the Chinese University of Hong Kong, Hong Kong, from 1997 to 1998. His research interests include chaos and information dynamics, signal processing, radar and communication technologies.
\end{IEEEbiography}

\end{document}